\documentclass{aa}
\usepackage{psfig}
\usepackage{natbib}
\usepackage{longtable}
\usepackage{graphicx}
\bibpunct{(}{)}{,}{a}{}{,} 

\def\la{\;
\raise0.3ex\hbox{$<$\kern-0.75em\raise-1.1ex\hbox{$\sim$}}\; }
\def\ga{\;
\raise0.3ex\hbox{$>$\kern-0.75em\raise-1.1ex\hbox{$\sim$}}\; }

\newcommand{\kms}{km~s$^{-1}$}

\newcommand{\cm}{cm$^{-2}$}
\newcommand{\cmm}{cm$^{-3}$}
\newcommand{\etal}{{et al.}}
\newcommand{\nhhh}{NH$_3$}
\newcommand{\nnhp}{N$_2$H$^+$}
\newcommand{\Dv}{$\Delta v$}
\newcommand{\Tmb}{$T_{\scriptscriptstyle \rm MB}$}
\newcommand{\VLSR}{$V_{\scriptscriptstyle \rm LSR}$}
\newcommand{\Tkin}{$T_{\rm kin}$}
\newcommand{\Trot}{$T_{\rm rot}$}
\newcommand{\Tex}{$T_{\rm ex}$}
\newcommand{\ta}{$\tau_{\scriptscriptstyle \rm 11}$}
\newcommand{\tb}{$\tau_{\scriptscriptstyle \rm 22}$}
\newcommand{\nHH}{$n_{\scriptscriptstyle {\rm H}_2}$}

\begin{document}

\title{Star-forming regions of the Aquila rift cloud complex
}
\subtitle{II. Turbulence in molecular cores probed by \nhhh\ emission}
\author{
S. A. Levshakov\inst{1,2,3}
\and
C. Henkel\inst{4,5}
\and
D. Reimers\inst{1}
\and
M. Wang\inst{6}
}
\institute{
Hamburger Sternwarte, Universit\"at Hamburg,
Gojenbergsweg 112, D-21029 Hamburg, Germany
\and
Ioffe Physical-Technical Institute,
Polytekhnicheskaya Str. 26, 194021 St.~Petersburg, Russia
\and
St.~Petersburg Electrotechnical University `LETI', Prof. Popov Str. 5,
197376 St.~Petersburg, Russia \\
\email{lev@astro.ioffe.rssi.ru}
\and
Max-Planck-Institut f\"ur Radioastronomie, Auf dem H\"ugel 69, D-53121 Bonn, Germany
\and
Astronomy Department, King Abdulaziz University, P.O.
Box 80203, Jeddah 21589, Saudi Arabia
\and
Purple Mountain Observatory, Key Laboratory of Radio Astronomy, 
Chinese Academy of Sciences, Nanjing 210008, P. R. China
}
\date{Received 00  ; Accepted 00}
\abstract
{}
{We intend to
derive statistical properties of stochastic gas motion inside the dense 
low mass star forming molecular cores
traced by \nhhh(1,1) and (2,2) emission lines.
}
{We use the spatial two-point autocorrelation (ACF) and structure functions 
calculated from maps of the radial velocity fields.
}
{The observed ammonia cores are characterized by complex intrinsic motions of stochastic nature.
The measured kinetic temperature ranges between 8.8~K and 15.1~K.
From \nhhh\ excitation temperatures of 3.5--7.3~K
we determine H$_2$ densities with typical values of \nHH\ $\sim(1-6)\times10^{4}$ \cmm. 
The ammonia abundance, $X=$ [\nhhh]/[H$_2$], varies from $2\times10^{-8}$ to $1.5\times10^{-7}$.
We find oscillating ACFs which eventually decay to zero with increasing lags
on scales of $0.04 \la \ell \la 0.5$ pc.
The current paradigm supposes that the star formation process is controlled by the interplay between
gravitation and turbulence, the latter preventing molecular cores from a rapid collapse due to
their own gravity. Thus, oscillating ACFs may indicate a damping of the developed turbulent
flows surrounding the dense but less turbulent core~-- a transition to dominating gravitational
forces and, hence, to gravitational collapse. 
}
{}
\keywords{ISM: clouds --- ISM: molecules --- ISM: kinematics and dynamics ---
Radio lines: ISM --- Line: profiles --- Techniques: spectroscopic
} 

\authorrunning{S. A. Levshakov \etal\ }

\titlerunning{Turbulence in molecular cores probed by \nhhh\ emission}

\maketitle

\section{Introduction}
\label{sect-1}

In the present study we continue our survey (Levshakov \etal\ 2013a, Paper~I)
of dense cores in Aquila in the \nhhh(1,1) and (2,2) lines. 
The targets selected are based on an extended CO survey with the Delingha 14-m telescope (Zuo \etal\ 2004),
the CO data of Kawamura \etal\ (1999, 2001), and
the optical data of Dobashi \etal\ (2005), limited in the latter case by areas with $A_V > 6$ mag.
More than 150 sources were detected in CO and $^{13}$CO at $\lambda \sim 3$~mm 
with the Delingha telescope (Wang 2012).
The angular resolution in these
observations was about $50''$ (FWHP, Full Width at Half Power) 
which is quite close to that of Effelsberg at K-band (FWHP $\simeq 40''$).
Subsequently, the NH$_3$ (1,1) and (2,2) emission lines were observed at Effelsberg with high
spectral resolutions. Channel widths were 
$\Delta_{\rm ch} = 0.015$, 0.038, and 0.077 km~s$^{-1}$ (Paper~I; Levshakov \etal\ 2013b).
Here we report on observations of newly discovered
three cores located in the star-forming region Serpens South, and,
for comparison, one core from the Cepheus Flare giant molecular cloud.

The current work is mainly focused on investigating the statistical characteristics of the spatial velocity field.
The most accurate information on the velocity field is 
provided by two-dimensional maps of
molecular lines toward a specific cloud,
with velocity providing a third dimension.
It follows that the observed profile of an optically thin spectral line
is a line of sight integral of emitted photons, and
that the measured line center, linewidth, and other parameters
result from a complex combination of such stochastic fields as 
gas density, velocity, temperature, and relative abundance.
To obtain the spatial distributions of physical parameters and kinematic
characteristics of the velocity field 
with the highest possible sensitivity, bright ammonia sources have been mapped.

Among the most sensitive species that trace the innermost parts of low-mass dense cores,
the ammonia molecule \nhhh\ plays a particular role:
\begin{itemize}
\item The observed inversion transitions \nhhh ($J,K$) = (1,1) and (2,2) at 23.7 GHz 
contain optically thin ($\tau \la 1$) and narrow ($FWHP \equiv \Delta v \la 1$  \kms)
spectral features.\footnote{Here $J$ is
the total angular momentum, and $K$ is the component of $J$ parallel to the molecular main axis.}.
\item In contrast to the carbon-chain molecules which 
represent an ``early'' chemical state because of later
freeze-out onto dust grains and gas phase reactions,
ammonia represents a ``more evolved'' chemical stage of evolution
and is typically observed in regions where the
density approaches $10^5$ \cmm\ (Gwenlan \etal\ 2000; Tafalla \etal\ 2004).
\item The rotational transitions with $\Delta K \ne 0$ are forbidden for ammonia, i.e.,
relative populations of ammonia rotational levels depend only on collisions and thus measure
the gas kinetic temperature \Tkin\ which is not affected by the presence of turbulent velocity components
(Juvela \etal\ 2012).
\item The inversion transitions have, in turn, magnetic hyperfine structure (hfs)
components of different intensities which 
provide a convenient tool to estimate the line optical depth $\tau_{\scriptscriptstyle J,K}$, the excitation
temperature \Tex, and the column densities $N_{\scriptscriptstyle J,K}$ 
(Ungerechts \etal\ 1980; Paper~I).
\item Detailed balance calculations 
between collisions and radiative processes
also provide a relation between
$n_{\scriptscriptstyle {\rm H}_2}$, \Tkin, and \Tex\ (Ho \& Townes 1983).
\item The comparison of the observable linewidths of the hfs components with their thermal widths
can be used to estimate the contribution of non-thermal (turbulent) velocity motion to the 
line broadening (Fuller \& Myers 1993).
\end{itemize}

Molecular cores~--- dense regions 
($n_{\scriptscriptstyle {\rm H}_2} \sim 10^4$ \cmm)
of low kinetic temperature (\Tkin $\sim 10$~K), small linear sizes ($L \sim 0.1$ pc)
and masses ($M \sim M_\odot$) embedded in molecular clouds 
($n_{\scriptscriptstyle {\rm H}_2} \sim 10^2$ \cmm, \Tkin $\sim 30$~K, $L \sim 10$ pc, $M \sim 10^2M_\odot$)~---
are believed to be the direct progenitors of individual 
low mass stars, binary systems, small stellar groups, or clusters.
Their physical properties are the key issue for gravitational fragmentation studies 
since low temperatures and relatively high gas densities favor gravitational contraction and collapse. 
Observations and numerical simulations show that the origin of molecular clouds is due to
the interplay of gravity and turbulence.
Turbulence is an effective mechanism creating 
overdensities in molecular clouds to initiate gravitational contraction
and to control star formation rates in interstellar clouds. 
Together with rotation, magnetic fields, and ambipolar diffusion, turbulence prevents molecular clouds
from a rapid collapse due to their own gravity. 

Turbulent (stochastic) motions in the cold
interstellar medium (ISM) are observed to be supersonic 
and are characterized 
by various physical processes linked nonlinearly over a wide range of scales 
from $\sim 150$ pc (the energy injection scale from supershells, e.g., Miesch \& Bally, 1994) 
down to the Kolmogorov dissipation
scale of $\sim 10^{14}$ cm~--- the smallest turbulent scale below which
viscous dissipation becomes dominant (e.g., Kritsuk \etal\ 2011). 
However, in spite of numerous theoretical and observational works 
(see the reviews of Elmegreen \& Scalo 2004;
McKee \& Ostriker 2007; Maury 2011; Klessen 2011; Crutcher 2012; Kennicutt \& Evans 2012; 
Hennebelle \& Falgarone 2012; Myers 2013),
real parameters of the small-scale stochastic motion in dense cores are still poorly known.
This makes the detailed study of molecular cores in ammonia lines particularly important.

The content of the paper is the following. Observations are described in Sect.~\ref{sect-2}.
Sect.~\ref{sect-3} deals with physical parameters of the observed dense cores.   
In Sect.~\ref{sect-4}, the velocity structure of the cores 
is analyzed by means of the autocorrelation and structure functions 
of the components of the velocity field parallel to the line of sight.
The results obtained are summarized in Sect.~\ref{sect-5}.
Appendix~A outlines the computational procedure used to analyze ammonia spectra.
The observed \nhhh\ spectra are presented in this appendix as well.

\section{Observations}
\label{sect-2}

The ammonia observations were carried out with the Effelsberg 100-m
telescope\footnote{The 100-m telescope at Effelsberg/Germany is operated
by the Max-Planck-Institut f{\"u}r Radioastronomie on behalf of the
Max-Planck-Gesellschaft (MPG).}
in two observing sessions between March 16 and 24, 2013, and in one session
on May 8 and 9, 2013 (Table~\ref{tbl-1}).
The measurements were performed in the frequency switching and position switching 
modes with the backend XFFTS (eXtended bandwidth FFTS) operating
with 100~MHz bandwidth and providing 32,768 channels
for each polarization. The resulting channel width was
$\Delta_{\rm ch} = 0.039$ \kms, but the true velocity resolution is 1.16 times
coarser (Klein \etal\ 2012).
The \nhhh\ lines were measured with a K-band high-electron mobility transistor (HEMT)
dual channel receiver, yielding spectra with a spatial resolution of $40''$ (FWHP) 
in two orthogonally oriented linear polarizations 
at the rest frequency of the $(J,K) = (1,1)$ (23694.495487 MHz) and (2,2) (23722.633644 MHz) lines
(Kukolich 1967). 
Averaging the emission from both
channels, the typical system temperature (receiver noise and atmosphere)
is 100\,K on a main beam brightness temperature scale.

The pointing was checked every hour by continuum cross scans
of nearby continuum sources. The pointing accuracy was better than $5''$. 
The spectral line data were calibrated by means
of continuum sources with known flux density. We mainly used G29.96--0.02 (Churchwell \etal\ 1990). 
With this calibration source a main beam
brightness temperature scale, \Tmb, could be established.
Since the main beam size ($40''$) is smaller than most core
radii ($>50''$) of our sources, the ammonia emission couples
well to the main beam and, thus, the \Tmb\ scale is appropriate.
Compensations for differences in elevation between the
calibrator and the source were $\la 20$\% and have not been taken
into account. Similar uncertainties of the main beam brightness
temperature were found from a comparison of spectra towards the
same position taken at different dates.

\section{Physical properties of molecular cores}
\label{sect-3}

Below we describe observations of the Cepheus molecular core \object{L1251C} and
the two sources \object{SS1} and \object{SS2}
(the latter consists of three separate objects labeled $A, B$ and $C$)
located in the star-forming region Serpens South which is associated with
the Aquila Rift. Together with \object{SS3} analyzed for the first time in ammonia lines in Paper~I, 
these sources form a triangle on the plane of 
the sky with sides $13.\!'$0 (SS1--SS3), $22.\!'3$ (SS3--SS2A), and 
$23.\!'7$ (SS2A--SS1).
\object{SS3} lies directly to the north of the central Serpens South protocluster (Gutermuth \etal\ 2008)
and coincides with a HC$_7$N(21-20) gas concentration at R.A. = 18$^{\rm h}$:29$^{\rm m}$:57$^{\rm s}$,
Dec. = $-1${\degr}:58{\arcmin}:55{\arcsec} (J2000) 
dubbed `1b' in Friesen \etal\ (2013).
\object{SS3} was recently observed in the HC$_5$N(9-8) emission line (Levshakov \etal\ 2013b), 
and in lines of \nnhp, H$^{13}$CO$^+$, NH$_2$D, SiO, HCO$^+$, CS, and HCN 
(Kirk \etal\ 2013; Tanaka \etal\ 2013).
However, \object{SS1} and \object{SS2} lie outside the regions mapped in these 
studies: the former in the west and the latter in the northwest 
direction to the Serpens South protocluster (see Fig.~1 in Friesen \etal\ 2013).

\begin{figure}[t]
\vspace{0.0cm}
\hspace{-0.6cm}\psfig{figure=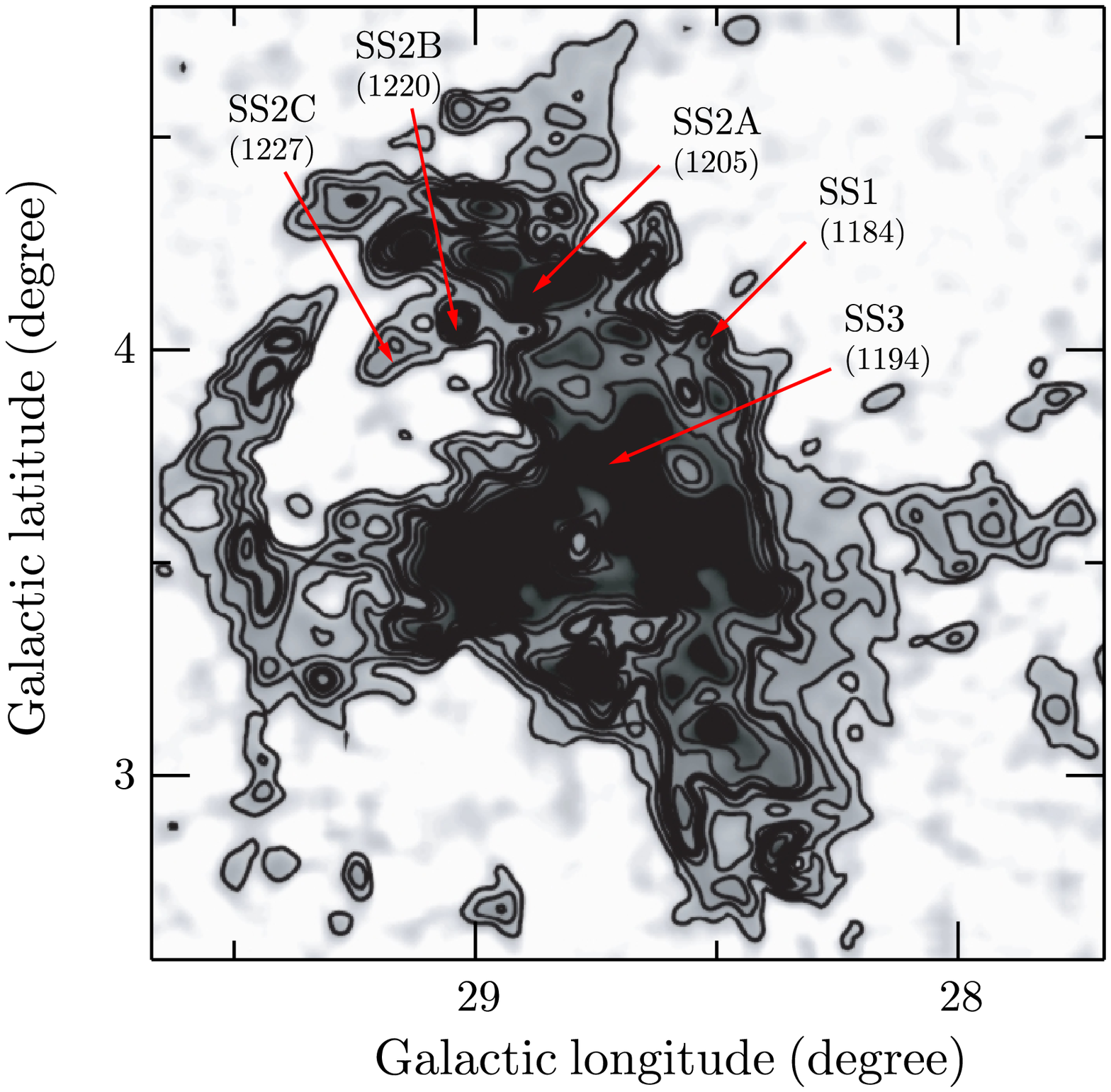,height=10.0cm,width=10.0cm}
\vspace{-0.7cm}
\caption[]{Dark cores from the 2 Micron All Sky Survey Point Source Catalog (2MASS PSC) by Dobashi (2011).
Contours start from $A^{\rm core}_V = 1.5$ mag, the increment is 1.5 mag, 
and thick lines are drawn at every four contours. 
The positions of the ammonia cores observed 
with the Effelsberg 100-m telescope
are marked by arrows (\object{SS3} was analyzed in Paper~I). 
Numbers in parentheses denote individual dust cores
compiled in Dobashi's catalog.
}
\label{fg1}
\end{figure}

Figure~\ref{fg1} shows dark cores detected in the Serpens South area  in
the 2 Micron All Sky Survey Point Source Catalog (2MASS PSC) by Dobashi (2011, D11).
Ammonia cores observed in the present work are marked by arrows. 
Numbers in parentheses denote dust cores compiled in Dobashi's catalog.
The following selection criteria were applied:
a continuous area having $A^{\rm core}_V \geq 1.5$ mag ($\sim 7.5\sigma$) with a single peak higher than
the boundary by $\geq 1$ mag ($\sim 5\sigma$) was defined as a candidate for a dark core.
The extinction value at the peak position for the Dobashi core \#1184
(the counterpart of \object{SS1}) is $A^{\rm core}_V = 9.6$ mag, 
for the three cores \#1205, \#1220, and \#1227 
(the counterparts of \object{SS2} $A, B$ and $C$) they are 13.5, 10.2, and 4.8 mag, respectively, and
it is 15.1 mag for the core \#1194 (the counterpart of \object{SS3}).
We note that the cores \#1184, 1220, and 1227 are labeled by the number \#279 in
the optical photographic database by Dobashi \etal\ (2005), and the cores
\#1205 and 1194 by \#279P22 and 279P7, respectively.
Surprisingly, the objects \object{SS2B} and \object{SS2C}
were not detected in CO lines from the Delingha observations.
Both of them were found serendipitously in our ammonia survey (see Sect.~\ref{sect-3-2}).

\subsection{\object{Serpens South 1} (\object{SS1})}
\label{sect-3-1}

\begin{figure}[t]
\vspace{0.0cm}
\hspace{-1.3cm}\psfig{figure=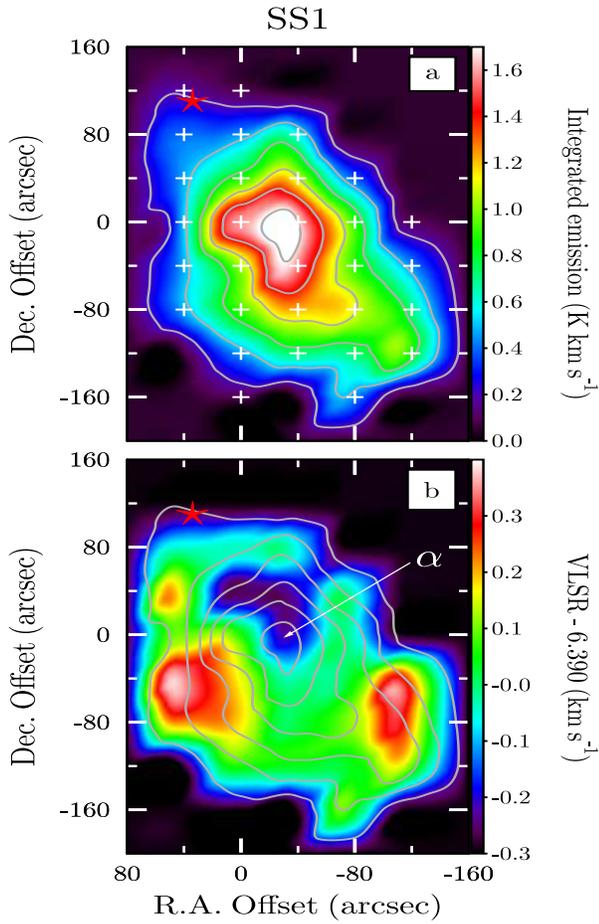,height=14.0cm,width=13.0cm}
\vspace{-0.7cm}
\caption[]{
({\bf a}): \nhhh(1,1) intensity map ($\int$\Tmb$dv$) of the molecular core \object{SS1}.
The starting point for the contour levels is 0.2 K~\kms, the increment is 0.3 K~\kms. 
Crosses mark measured positions, the red star indicates
the location of the IR source IRAS 18265--0205 (Beichman \etal\ 1988).
({\bf b}): \nhhh(1,1) radial velocity field (color map)
shown after subtracting the mean radial velocity $\langle$\VLSR$\rangle$ = 6.390 \kms.
The \nhhh\ peak is labeled as in Table~\ref{tbl-1}.
The (0,0) map position is R.A. = 18:29:09.6, Dec. = $-02$:05:40 (J2000).
}
\label{fg2}
\end{figure}

Our \nhhh\ observations cover the whole molecular core \object{SS1} and consist of 30
spectra obtained at the positions marked by crosses in Fig.~\ref{fg2}{\bf a}.
We adopt $D = 203$ pc as the distance to the source (Paper~I).   
The angular size of the major (NE-SW) and minor (SE-NW) axis\footnote{The core size is defined
at the half-maximum (50\%) contour of the integrated intensity map of the corresponding emission line.}
is  $175'' \times 125''$. 
This corresponds to the linear dimensions $a \times b \approx 0.17$~pc $\times 0.12$~pc 
and to the mean diameter $L = \sqrt{a \cdot b} \approx 0.14$ pc.
Dust emission at 2~$\mu$m was detected from 
core \#1184 (D11)~--- the counterpart of the source \object{SS1}. 
The dust emission is seen from an area $S = 79.8$ arcmin$^2$ (D11),
corresponding to an extent of $\sim 536''$, which is about four times the ammonia emitting region.
The core is situated slightly above the Galactic plane 
($b \simeq 4$\degr\ corresponding to $\sim$15 pc)\footnote{This implies that the core is in total
$\sim$35 pc above the Galactic midplane, if the Sun's distance is $\sim$20 pc (Humphreys \etal\ 1995).}
and consists of a dense core labeled $\alpha$ in Fig.~\ref{fg2}{\bf b}
and a weak filament pointing south-west.
An infrared source IRAS 18265--0205 (Beichman \etal\ 1988) located at the 
north-eastern edge of the core is marked by a red star in Fig.~\ref{fg2}.
It is not clear if this IR source is physically connected with the molecular core
since no related perturbations or systematic flows are seen in Fig.~\ref{fg2}{\bf b}.

The extended central region of \object{SS1} covers three adjacent positions of 
similar brightness temperature \Tmb\ $\sim 1$~K (see Table~\ref{tbl-1}) which
are not resolved at our angular resolution of $40''$ (the corresponding spatial resolution is $\sim 0.04$ pc).
However, the positions show essentially different radial velocities, \VLSR\ =  
6.41, 6.139, and 6.32 \kms, and linewidths, \Dv\ = 0.69, 0.54, and 0.95 \kms,
that may point to the existence of substructure at scales $< 0.04$ pc.
Formally, the position $(-40'',0'')$ with the narrowest line and a slightly higher line temperature 
is referred to as  the ``$\alpha$ peak''.

The model fit to the observed \nhhh(1,1) and (2,2) spectra is shown in 
Fig.~\ref{fg1a} and the corresponding model parameters are given
in Table~\ref{tbl-1a}.
Toward the $(0'',0'')$ position, we measured 
a kinetic temperature of \Tkin\ $\sim 15$~K~--- 
rather high for dense molecular cores~--- 
and a typical gas density of \nHH\ $\sim (2-3)\times10^4$ \cmm.

At the $\alpha$ peak, the total ammonia column density $N_{\rm tot} = 1.8\times10^{14}$ \cm\ 
and the gas density \nHH\ $= 1.8\times10^4$ \cmm\ (Table~\ref{tbl-1a})
correspond to the ammonia abundance ratio $X=$ [\nhhh]/[H$_2$] $= 2\times10^{-8}$,
which is in line with other sources (e.g., Dunham \etal\ 2011).

The measured linewidth at the $\alpha$ peak
\Dv\ = 0.54 \kms\ and the kinetic temperature \Tkin\ = 13.6~K (Table~\ref{tbl-1a})
give a thermal velocity $v_{\rm th} = 0.11$ \kms\ [Eq.~(\ref{Eq5a})],
and a turbulent velocity dispersion $\sigma_{\rm turb} = 0.22$ \kms\ [Eq.~(\ref{Eq6a})].
The dispersion of the line-of-sight velocity component
of bulk motions is comparable to
the sound speed  $c_s \simeq 0.22$ \kms\  [Eq.~(\ref{Eq7a})], i.e.,
non-thermal motions are transonic with
the Mach number ${\cal M}_s = 1\pm0.1$ [Eqs.~(\ref{Eq8a}), (\ref{Eq08a})]. 
However, supersonic motions with local 
${\cal M}_s = 1.2\pm0.1$ and $1.8\pm0.1$ are observed towards the two other peaks at offsets
$(0'',0'')$ and $(-40'',-40'')$, respectively.
The transition between transonic and supersonic gas motions is observed over a beam width ($0.04$ pc). 
For example, the change in the velocity dispersion, $d\sigma_{\rm turb}/dr$,
between the $\alpha$ peak and the offset $(-40'',-40'')$ is
$d\sigma_{\rm turb}/dr \ga 4$ km~s$^{-1}$~pc$^{-1}$, 
which is comparable to the change $d\sigma_{\rm turb}/dr \approx 3$ km~s$^{-1}$~pc$^{-1}$
observed towards the B5 region in Perseus (Pineda \etal\ 2010).
Such a sharp transition~--- the increase of the velocity dispersion by a factor 1.5--2 in less than 0.04 pc~---
is also observed in other cores mapped in the \nhhh(1,1) line at high angular resolutions
(see, e.g., the review by Andr\'e \etal\ 2013).

The gas density, \nHH, determined for the
three central condensations and the ammonia linewidths, $\Delta v$, (Table~\ref{tbl-1a})
can be used to estimate a core mass.
Using the mean gas density
$n_{\scriptscriptstyle {\rm H}_2} = 2.4\times10^4$ \cmm, and the core radius $R \sim 0.07$ pc, and
assuming spherical geometry, we find a core mass
$M \sim 2M_\odot$ (the mean molecular weight is 2.8). 
On the other hand,
the virial mass [Eq.~(\ref{Eq9a})] calculated from the weighted mean linewidth  $\Delta v = 0.77$ \kms\
and accounting for the dominating non-thermal line broadening
is $M_{\rm vir} \sim 10M_\odot$. 
The observed difference in
the masses could be due to deviations from a uniform gas density
distribution and a core ellipticity.

\subsection{\object{Serpens South 2} (\object{SS2})}
\label{sect-3-2}

\begin{figure*}[t]
\vspace{-3.0cm}
\hspace{-0.5cm}\psfig{figure=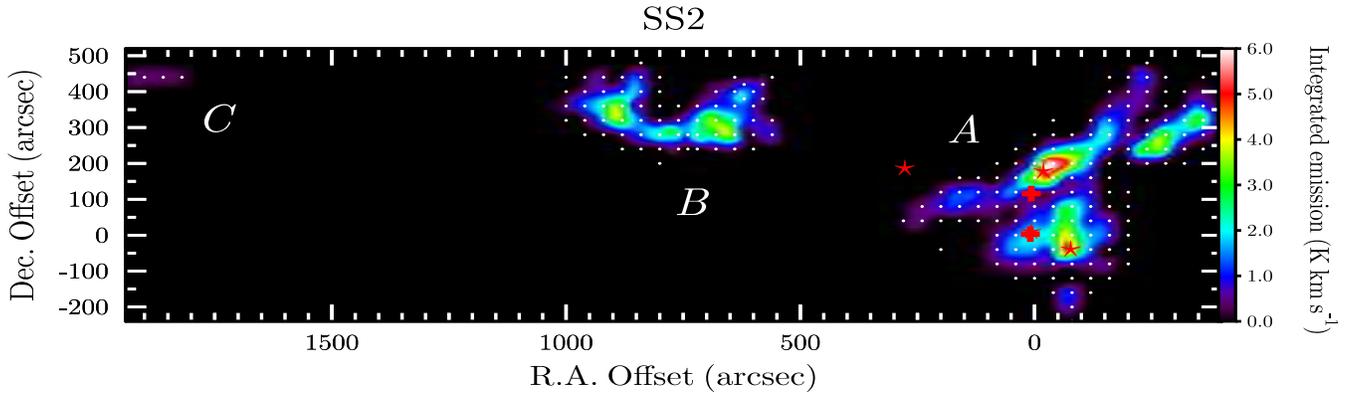,height=22.0cm,width=19.0cm}
\vspace{-8.5cm}
\caption[]{The integrated \nhhh(1,1) emission
($\int$\Tmb$dv$) of the molecular core \object{SS2}
(source $A$) and two serendipitously detected cores (sources $B$ and $C$).
White dots mark measured positions, crosses~---
the location of the IR sources IRAS 18264--0143 and IRAS 18264--0142
(Beichman \etal\ 1988), and stars~--- 
Class 0 protostars (Maury \etal\ 2011).
The (0,0) map position is 
R.A. = 18:29:05.1, Dec. = $-01$:42:00 (J2000).
}
\label{fg3}
\end{figure*}

Starting observations of the main source \object{SS2A} in the position switching mode, we detected 
serendipitously two other ammonia compact cores which we refer to as \object{SS2B}
and \object{SS2C} (Fig.~\ref{fg3}).
Their spectral lines are shifted with respect to those of \object{SS2A}: 
$V_{\scriptscriptstyle {\rm LSR},A} = 7.4$ \kms,
$V_{\scriptscriptstyle {\rm LSR},B} = 8.2$ \kms, and
$V_{\scriptscriptstyle {\rm LSR},C} = 8.5$ \kms\ (Table~\ref{tbl-1}).
The source \object{SS2B} was completely mapped down to its edge, whereas we did not have enough
time to investigate the area around \object{SS2C}. For this reason only three \nhhh(1,1)
spectra of \object{SS2C} can be shown in Fig.~\ref{fg2a}.

The strongest ammonia emission measured at \VLSR\ $= 8.5$ \kms\ towards \object{SS2C}  
is one of the narrowest lines (\Dv\ $=0.25\pm0.03$ \kms) observed in our survey (see Table~\ref{tbl-1}).
If we assume for this source a kinetic temperature \Tkin\ $\sim 10$~K, then the estimated 
$\sigma_{\rm turb} \sim 0.08$ \kms\ would imply gas moving at subsonic velocities with
${\cal M}_s \approx 0.4\pm0.2$.  
The dust core \#1227 from Dobashi's catalog (D11), a counterpart of the source \object{SS2C}, 
occupies a region of 65.8 arcmin$^2$ with a linear size of about $490''$. 
This is the only information available for this source at the moment.
In the following subsections we will concentrate on our results obtained for \object{SS2A} and \object{SS2B}.

\subsubsection{\object{SS2A} }
\label{sect-3-2-1}

\begin{figure}[t]
\vspace{0.0cm}
\hspace{-1.0cm}\psfig{figure=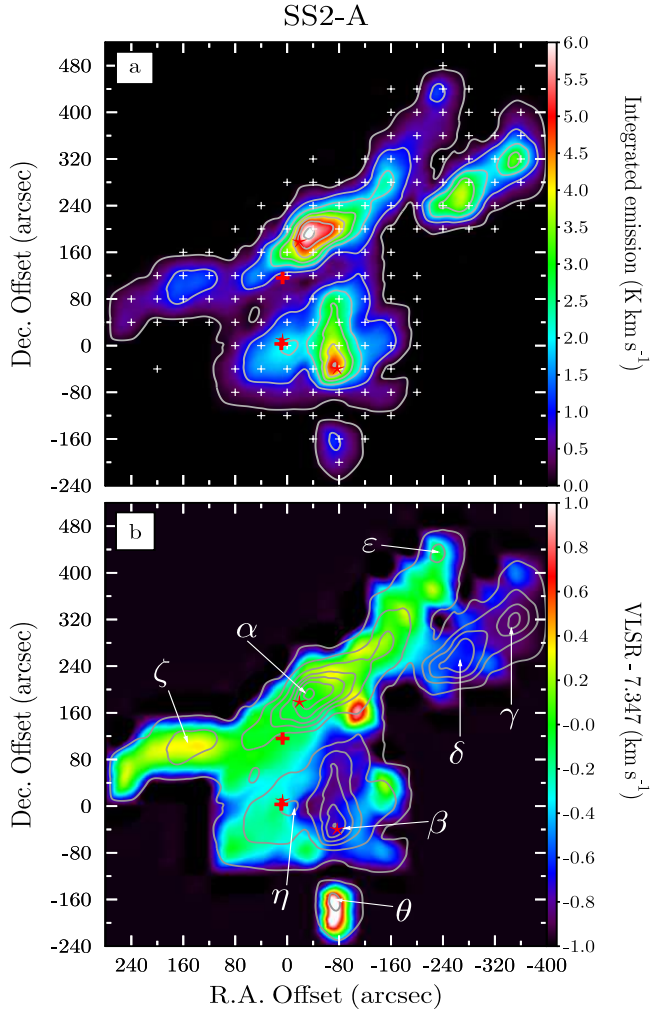,height=14.0cm,width=11.0cm}
\vspace{-0.7cm}
\caption[]{A zoomed portion of the map shown in Fig.~\ref{fg3}.
({\bf a}): \nhhh(1,1) intensity map of \object{SS2A}.
The starting point for the contour levels is 0.1 K~\kms, the increment
is 0.9 K~\kms\ between the first two and 1.0 K~\kms\ between the other contour levels.
White crosses mark measured positions, red crosses~---
the location of the IR sources IRAS 18264--0143 and IRAS 18264--0142 (Beichman \etal\ 1988),
and red stars~--- Class 0 protostars (Maury \etal\ 2011).
({\bf b}): \nhhh(1,1) radial velocity field (color map)
shown after subtracting the mean radial velocity $\langle$\VLSR$\rangle$ = 7.347 \kms.
The ammonia intensity peaks are labeled as in Table~\ref{tbl-1}.
The origin of the map position is as in Fig.~\ref{fg3}.
}
\label{fg4}
\end{figure}

The source \object{SS2A} is an example of a complex molecular core 
consisting of at least eight ammonia peaks (Fig.~\ref{fg4}).
Individual \nhhh\ peaks are labeled $\alpha$ through $\theta$ in order of decreasing
peak line brightness temperature \Tmb\ (Table~\ref{tbl-1}).
The ammonia emission is observed from an elongated irregular structure in
the NW-SE direction over an area of $\approx 600'' \times 400''$ 
($\approx 0.6$ pc $\times$ 0.4 pc). 
The associated dust core \#1205 extends over an area of 170.6 arcmin$^2$ (D11) and has
a linear size of about $780''$.
The core harbors a few IR sources and protostars
marked by red crosses and red stars in Fig.~\ref{fg4}.
The $\eta$ peak coincides with the class~0 protostar MM2 which is a 
condensation of several infrared sources (Maury \etal\ 2011).
The measured positions are marked by white crosses.

Figure~\ref{fg4}{\bf b} shows a rather high amplitude of the radial velocity
fluctuations, $\delta V_{\scriptscriptstyle \rm LSR} \simeq \pm 1$ \kms, as compared with 
the velocity fields of \object{SS1} 
(Fig.~\ref{fg2}) and \object{SS2B} (Fig.~\ref{fg5}{\bf b}), where the observed amplitude is
$\delta V_{\scriptscriptstyle \rm LSR} \simeq \pm 0.3$ \kms.
High velocity perturbations in \object{SS2A} resemble a velocity pattern measured towards
the core \object{Do279P12} where $\delta V_{\scriptscriptstyle \rm LSR} \simeq \pm 1$ \kms\ and 
the \nhhh(1,1) and (2,2) lines are split 
at some positions into two components (Paper~I).

A similar picture of a split but not completely resolved line profile  showing a redward asymmetry
is observed towards \object{SS2A}. 
The ammonia profiles depicted in 
Fig.~\ref{fg3a} at offsets $(-40'',0'')$, $(-80'',0'')$, and $(-80'',40'')$
cannot be described by a single component model but require a second component shifted with respect
to the first one by 0.8, 0.7, and 0.9 \kms, respectively (see Table~\ref{tbl-1a}).
On the other hand, the ammonia magnetic hfs components are optically thin ($\tau < 1$) at these velocity offsets.
Therefore, the observed asymmetries must
be due to kinematics rather than radiative transfer effects  
expected at high optical depths\footnote{Saturated emission lines may have 
``red shoulders'' due to self-absorption that are indicative of collapse (e.g., Williams \etal\ 2006).}.

The \nhhh\ linewidths range in a wide interval between \Dv\ = 0.22 \kms\ and 1.04 \kms\ (Table~\ref{tbl-1}).
Unfortunately, the second inversion line \nhhh(2,2) was not detected at all ammonia peaks listed in Table~\ref{tbl-1},
and we were not able to estimate directly the kinetic temperature and, thus, the
contribution of turbulent and thermal 
motions to the line broadening at all offsets.
However, at 15 positions listed in Table~\ref{tbl-1a} we observed both the (1,1) and (2,2) lines and, thus,
the values of $\sigma_{\rm turb}$ and $v_{\rm th}$ can be calculated directly.
The maximum dispersions of $\sigma_{\rm turb} = 0.22$ \kms\ (\Tkin\ = 14.2~K) and 
$\sigma_{\rm turb} = 0.24$ \kms\ (\Tkin\ = 10.3~K) are found at offsets 
$(-40'',160'')$ ($\Delta v = 0.54\pm0.01$ \kms)
and $(-80'',0'')$ ($\Delta v = 0.590\pm0.007$ \kms) 
where we observe transonic and supersonic motions with ${\cal M}_s = 1.0\pm0.1$ and $1.3\pm0.1$, respectively. 
However, the latter position reveals an asymmetry of the ammonia 
magnetic hfs components with a formally deconvolved second
subcomponent having a wider linewidth, \Dv\ = $1.26\pm0.04$ \kms\   
(Table~\ref{tbl-1a}). Assuming the same \Tkin\ as measured
for the main subcomponent, one obtains $\sigma_{\rm turb} \sim 0.53$ \kms\ and ${\cal M}_s = 2.8\pm0.1$. 
The velocity dispersion of the weaker subcomponents 
at two other positions with asymmetric \nhhh\ lines is of the same order of magnitude, presenting strongly supersonic
non-thermal motions at small scales. The regions with asymmetric profiles are 
probably affected by the closely located IR source IRAS 18264--0143
and protostars at the $\beta$ and $\eta$ peaks (see Fig.~\ref{fg4}).   
At all other positions, the dispersions of the random velocity
component and the corresponding Mach numbers lie in the intervals
0.05 \kms\ $\leq \sigma_{\rm turb} \leq 0.19$ \kms\ and 
$0.2 \leq {\cal M}_s \leq 0.9$ which favors subsonic and transonic motions.
Here again, as in the case of \object{SS1}, we observe the transition between subsonic and supersonic 
gas motions occurring over a very short distance $\ell < 0.04$ pc. 
For example, the narrower component of the \nhhh(1,1) line at offset $(-80'',0'')$ 
corresponds to ${\cal M}_s = 1.3\pm0.1$,  
whereas at two neighboring positions
$(-80'',40'')$ ($\Delta v = 0.28\pm0.01$ \kms) and $(-40'',0'')$ ($\Delta v = 0.26\pm0.01$ \kms), 
the local Mach numbers are $0.5\pm0.1$ and $0.4\pm0.2$, 
which implies $d\sigma_{\rm turb}/dr \ga 4$ km~s$^{-1}$~pc$^{-1}$.

The deconvolved angular size of the $\alpha$ and $\beta$ peaks is $124'' \times 45''$ (0.12 pc $\times 0.04$ pc)
that gives the mean diameter $\sim 0.07$ pc.
The gas densities were determined only at the $\alpha, \beta, \gamma$, and $\delta$ peaks
(Table~\ref{tbl-1a}). Following the same procedure as in Sect.~\ref{sect-3-1},
the masses of the $\alpha$ and $\beta$ peaks are $M_\alpha \sim M_\beta = 0.3M_\odot$. 
The $\gamma$ and $\delta$ peaks are unresolved, i.e., their diameter is less than 0.04 pc.
This give an estimate $M_\gamma, M_\delta < 0.04M_\odot$. The virial masses of these peaks
are $M_{\rm vir} = 2M_\odot (\alpha)$, and $1.5M_\odot (\beta)$, and 
$M_{\rm vir} < 0.3M_\odot (\gamma)$, and $< 0.5M_\odot (\delta)$.
The gas densities at the unresolved $\varepsilon, \zeta, \theta$, and $\eta$ peaks
are not known, so that we can estimate only virial masses:
$M_{\rm vir} < 0.2M_\odot (\varepsilon), < 0.3M_\odot (\zeta), < 1.1M_\odot (\theta)$, and
$< 5.4M_\odot (\eta)$.
 
At the $\alpha$ peak, the total ammonia column density $N$(\nhhh) = $9.1\times10^{14}$ \cm,
and the gas volume density \nHH\ = $2.7\times10^4$ \cmm\ give the abundance ratio
[\nhhh]/[H$_2$] $\sim 1.5\times10^{-7}$.

\subsubsection{\object{SS2B} }
\label{sect-3-2-2}

\begin{figure}[t]
\vspace{-3.0cm}
\hspace{-1.0cm}\psfig{figure=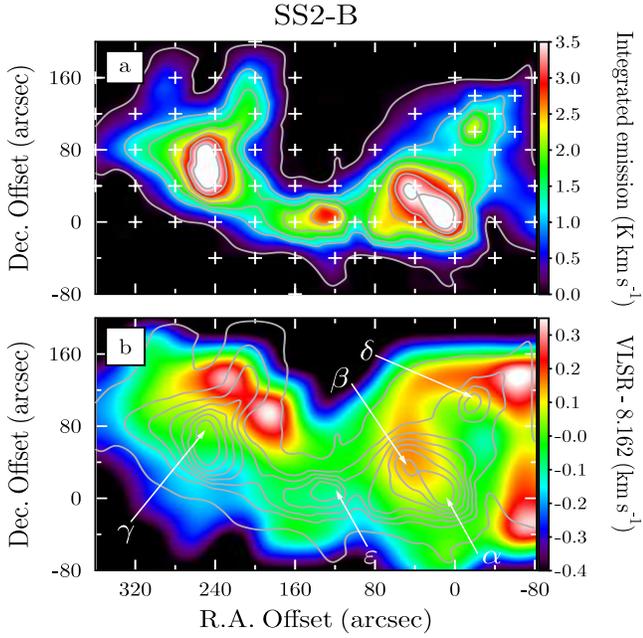,height=14.0cm,width=11.0cm}
\vspace{-1.5cm}
\caption[]{A zoomed portion of the map shown in Fig.~\ref{fg3}.
({\bf a}): \nhhh(1,1) intensity map ($\int$\Tmb$dv$) of \object{SS2B}.
The starting point for the contour levels is 0.1 K~\kms, the increment
is 0.9 K~\kms\ between the first two and 0.5 K~\kms\ between the other contour
levels. Crosses mark measured positions.
({\bf b}): \nhhh(1,1) radial velocity field (color map)
shown after subtracting the mean radial velocity $\langle$\VLSR$\rangle$ = 8.162 \kms.
The ammonia intensity peaks are labeled as in Table~\ref{tbl-1}.
The (0,0) map position is 
R.A. = 18:29:47.8, Dec. = $-01$:37:20 (J2000).
}
\label{fg5}
\end{figure}

The serendipitously detected source \object{SS2B} is 
a presumably starless core\footnote{The starless nature of \object{SS2B} is deduced from the lack
of any {\it IRAS} point sources within the boundaries of the core.}
curved with a ''C'' shape.
The surface area 82.8 arcmin$^2$ at 2 $\mu$m of the 
counterpart \#1220 forms approximately a square with side lengths
$\sim 550''$ (Fig.~35, panel 43-1 in D11). 
The dust core covers completely the ammonia
emission having a largest extent of $\la 400''$
and showing five ammonia peaks (Fig.~\ref{fg5}).
The peaks are labeled $\alpha$ through $\varepsilon$ in order of decreasing
peak line brightness temperature \Tmb\ (Table~\ref{tbl-1}).
The $\alpha$ and $\beta$ peaks in the west are just two marginal points along an elongated condensation on
the integrated intensity map around the origin $(0'',0'')$ (Fig.~\ref{fg5}{\bf a}).
They have practically identical physical parameters (Table~\ref{tbl-1a}) but slightly
different radial velocities \VLSR\ = 8.19 and 8.32 \kms,
linewidths \Dv\ = 0.259 and 0.287 \kms, and non-thermal velocity dispersions
$\sigma_{\rm turb} = 0.087$ and 0.103 \kms\ for the $\alpha$ and $\beta$ peaks, respectively.
The $\gamma$-peak located at the eastern side of the C-shaped core 
shows with  \Tmb\ $= 2.2$~K the same brightness temperature 
as the previously mentioned peaks, whereas its linewidth is 1.5 times larger,
\Dv\ = 0.43 \kms, and $\sigma_{\rm turb} = 0.169$ \kms.
The observed linewidths at the $\delta$ and $\varepsilon$ peaks are as narrow as those towards
the $\alpha$ and $\beta$ peaks.

The kinetic temperature measured at nine positions towards \object{SS2B} is on the average lower than that
in the two other sources \object{SS1} and \object{SS2A}. Moreover, at the offset $(280'',80'')$, nearby
the $\gamma$ peak, we detected the lowest kinetic temperature \Tkin\ = 8.8~K among 
\Tkin\ listed in Table~\ref{tbl-1a}.
The measured values of \Tkin\ were used to estimate 
the dispersion of non-thermal gas motions and the Mach numbers from the observed linewidths. 
Calculations reveal transonic velocities with 
${\cal M}_s = 0.9\pm0.1$ and $0.8\pm0.1$ at the $\gamma$ peak ($\Delta v = 0.426\pm0.006$ \kms)
and at the offset $(80'',40'')$ ($\Delta v = 0.44\pm0.01$ \kms) located 
between the $\varepsilon$ and $\beta$ peaks (see Fig.~\ref{fg5}{\bf b}).
At all other positions we observe clearly subsonic motions with $0.4 \la {\cal M}_s \la 0.7$.
The positions with the transonic velocities demonstrate a sharp change of the non-thermal
velocity dispersion over a distance $\ell \la 0.04$ pc.  
Namely,
$d\sigma_{\rm turb}/dr \ga 2$ km~s$^{-1}$~pc$^{-1}$
between the $\beta$ peak and the offset $(80'',40'')$, and between
the $\gamma$ peak and the narrower \nhhh\ component at $(280'',80'')$.

As in the case of asymmetric ammonia profiles discussed in the previous subsection,
here we detected an asymmetric line at the 
already mentioned offset $(280'',80'')$, this time showing
a blueward asymmetry: this subcomponent is shifted by $-0.35$ \kms\
with respect to the first one (Fig.~\ref{fg4a}). Both subcomponents are optically thin:
for the strongest magnetic hfs component 
($F'_1,F'- F_1,F = 2,5/2 - 2,5/2$) the optical depths equal to
$\tau = 0.3$ and 0.4, respectively.
Since $\tau$ is well below unity, the observed asymmetry must be caused by kinematics
rather than radiative transfer effects. 

The angular size of the combined $\alpha + \beta$ peak is $120'' \times 120''$
(0.12 pc $\times$ 0.12 pc, $R \sim 0.06$ pc).
Applying the same procedure as in Sect.~\ref{sect-3-1}, we find the 
weighted mean linewidth $\Delta v = 0.28$ \kms, and the mean gas density \nHH\ = $2.2\times10^4$ \cmm.
This gives $M = 1.4M_\odot$ and the virial mass $M_{\rm vir} = 1.1M_\odot$.
The deconvolved angular size of the $\gamma$ peak $70'' \times 45''$ 
(0.07 pc $\times$ 0.04 pc, $R \sim 0.03$ pc) gives 
$M = 0.6M_\odot$ and $M_{\rm vir} \sim 1.8M_\odot$. 
The structure of the $\delta$ and $\varepsilon$ peaks is spatially unresolved
($R < 0.02$ pc), and their gas densities are not known. 
In this case only upper limits on the virial masses can be estimated:
$M_{\rm vir} < 0.3M_\odot (\delta)$, and $< 0.4M_\odot (\varepsilon)$.

The ammonia abundance at the resolved $\alpha +\beta$ peak is 
$X_{\alpha, \beta} \sim 9\times10^{-8}$ if the beam filling factor $\eta = 1$.

\subsection{\object{L1251C} }
\label{sect-3-3}

\object{L1251} is a small E-W elongated dark core of the opacity class 5 
(the second highest, see Lynds 1962)
which is located at a distance $D = 300\pm50$ pc (Kun \& Prusti 1993).
With the coordinates
$\ell \approx 115$\degr, $b \approx 15$\degr,
its distance from the Galactic midplane is about 100 pc.
It is a part of the Cepheus Flare giant molecular cloud complex (Lebrun 1986).
Five C$^{18}$O cores embedded in the $^{13}$CO cloud were revealed by Sato \etal\ (1994).
The northern core \object{L1251C} is one of them. 
The measured width of the C$^{18}$O(1--0) line toward the peak,
observed with a beam size of $2.\!'7$, 
is \Dv\ = 1.7 \kms, the radial
velocity is \VLSR\ = $-4.5$ \kms, and the core size is $1.2\times0.56$ pc
if $D = 300$ pc (Sato \etal\ 1994).
The 2~$\mu$m emission associated with \object{L1251C} was detected towards
the dust core \#3553 from Dobashi's catalog (D11). According to D11, 
$A^{\rm core}_V = 8.4$ mag at the position of the 2~$\mu$m peak, and 
the surface area 
$S = 233.12$ arcmin$^2$ (angular size $\sim 916''$).

Eight ammonia cores were detected in a map of 0.12 square degrees
sampled at intervals of $80''$ (2 times the FWHP)
with the Effelsberg 100-m telescope (T\'oth \& Walmsley 1996).
The observed parameters for the northern core \object{L1251C}
are the following:
$N$(\nhhh)$_{\rm max} = 8.06\times10^{14}$ \cm, 
\nHH$_{\rm max} = 2.6\times10^4$ \cmm, \VLSR\ = $-4.79$ \kms, and \Dv\ = 0.30 \kms\
(T\'oth \& Walmsley 1996). 

\begin{figure}
\hspace{-0.5cm}\psfig{figure=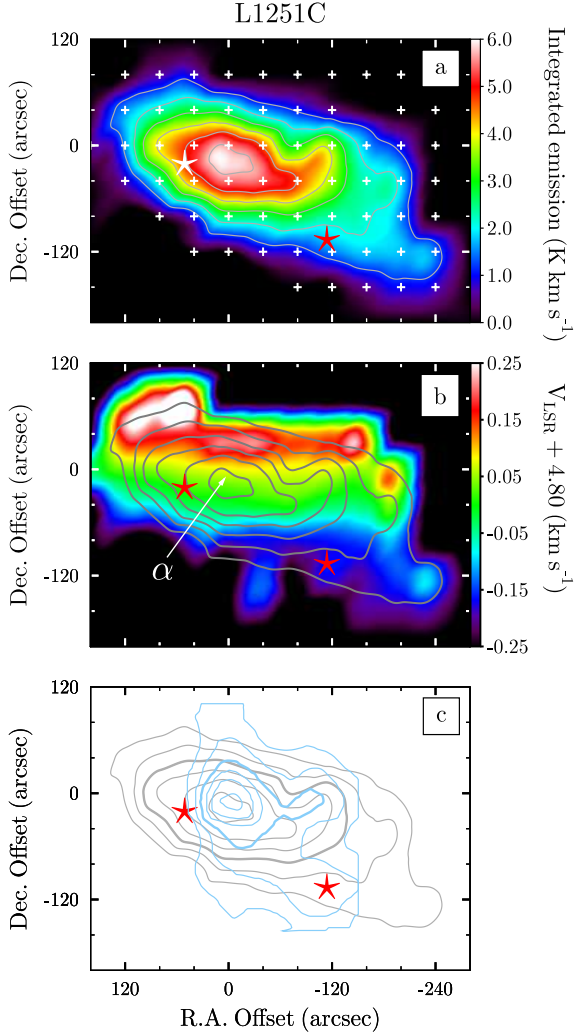,height=16.0cm,width=11.0cm}
\vspace{-2.0cm}
\caption[]{
({\bf a}): \nhhh(1,1) intensity map ($\int$\Tmb$dv$) of \object{L1251C}.
The starting point for the contour levels is 1.2 K~\kms, the increment is 0.9 K~\kms.
The crosses mark measured positions. The stars indicate the location of the IR sources detected by $IRAS$.
({\bf b}): \nhhh(1,1) radial velocity field (color map)
shown after subtracting the mean radial
velocity $\langle V_{\scriptscriptstyle\rm LSR} \rangle = -4.80$ \kms.
The \nhhh\ peak is labeled as in Table~\ref{tbl-1}.
The (0,0) map position is R.A. = 22:35:53.6, Dec. = 75:18:55 (J2000).
({\bf c}): The grey and blue contours show, respectively, the \nhhh(1,1) and N$_2$H$^+$(1--0)
intensities integrated over all hyperfine components. 
The thick contours denote the half-maximum (50\%) level which defines the core size in the corresponding emission line.
The N$_2$H$^+$(1--0) data are taken from Fig.~2 of Caselli \etal\ (2002),
where the contour levels are 0.4, 0.7, 1.0, 1.2, 1.5, and 1.8 K~\kms, and the peak intensity is
1.91$\pm0.12$ K~\kms.
}
\label{fg6}
\end{figure}

We mapped the core \object{L1251C} in the \nhhh(1,1) and (2,2) lines with a spacing of 
$40''$ towards 60 positions marked by crosses in Fig.~\ref{fg6}{\bf a}.  
This confirms that the \nhhh\ emission arises from a very compact central region 
marked by $\alpha$ in Fig.~\ref{fg6}{\bf b}. 
The third contour level in Fig.~\ref{fg6} corresponding
to the half-peak of the integrated emission, $\int$\Tmb$dv$, 
defines the core size in the \nhhh(1,1) emission line: $a \times b \approx 240'' \times 100''$,
i.e., $a \times b \approx 0.35$~pc $\times$ 0.15~pc if $D = 300$ pc.

The results of the fits to the \nhhh(1,1) and (2,2) spectra 
are shown by red curves in Fig.~\ref{fg5a} and the estimated physical parameters
are given in Table~\ref{tbl-2a}.
The excitation temperature \Tex\ is calculated from the (1,1) transition assuming a beam-filling
factor $\eta = 1$ since both the major and minor axes of the core exceed the angular resolution.
We calculated \Trot\ and \Tkin\ at all offsets where both
the (1,1) and (2,2) transitions were observed (for details, see Appendix~A in Paper~I).
At the core center, offset ($0'',0''$), the derived parameters $N_{\rm tot}$, \Tkin,
and \Dv\ are in concordance with those of T\'oth \& Walmsley (1996), but we
measure a two times higher gas density, \nHH\ $= 5.9\times10^4$ \cmm.

The total ammonia column density of $N_{\rm tot} = 8\times10^{14}$ \cm\ 
gives the abundance ratio $X=$ [\nhhh]/[H$_2$] $= 2\times10^{-8}$
at the core center of the mean size $L = \sqrt{a \cdot b} \approx 0.2$ pc.
The same ammonia abundance was found towards the core \object{SS1} (Sect.~\ref{sect-3-1}).

The completely resolved ammonia spectra can be used
to estimate the contribution of the non-thermal (turbulent) motions to the line broadening.
The measured linewidth \Dv\ = $0.290\pm0.007$ \kms\  
and the kinetic temperature \Tkin\ = 10.5~K
at the core center give a thermal velocity  $v_{\rm th} = 0.10$ \kms,
a dispersion of the turbulent motions
$\sigma_{\rm turb} = 0.10$ \kms,
a sound speed $c_s = 0.19$ \kms,
and a local Mach number ${\cal M}_s = 0.5\pm0.1$ 
[see Eqs.~(\ref{Eq5a})--(\ref{Eq08a}) in Appendix~A]. 
Subsonic motions of the non-thermal velocity component 
are also observed at other positions where ${\cal M}_s$ ranges from 0.3 to 0.6.
Thus the core as a whole gives an example of moderately quiet bulk motions 
with the ratio of the non-thermal velocity dispersion to the thermal
velocity of order unity.

The velocity map of L1251C is shown in Fig.~\ref{fg6}{\bf b}.
All along the major axis (E-W direction, $\Delta\delta = -40''$)
the radial velocity is close to its mean value $\langle$\VLSR$\rangle = -4.80$ \kms.
In this region, the dispersion of bulk motions is subsonic and roughly constant,
$\sigma_{\rm turb} = 0.12\pm0.1$ \kms, that corresponds to ${\cal M}_s \simeq 0.6$.
Such regions are usually called `coherent' (Goodman \etal\ 1998; Caselli \etal\ 2002;
Pineda \etal\ 2010)\footnote{Similar coherent zones were observed in the molecular cores studied in Paper~I.}.
However, the radial velocity monotonically changes by $\pm0.2$ \kms\ towards the northern and southern edges
of the core.
As they are attributed to core rotation, these systematic changes
in \VLSR\ correspond to an angular velocity $\dot{\phi} \approx 7\times10^{-14}$ s$^{-1}$.

For a self-gravitating rigidly rotating sphere of constant
density $\rho$, the ratio of rotational to
gravitational binding energy is
\begin{equation}
p = \frac{\dot{\phi}^2}{4\pi G \rho} = 2.55\times10^{-3}{\dot{\phi}}^2_{-14}/n_4, 
\label{Eq1b}
\end{equation}
where $G$ is the gravitational constant,
${\dot{\phi}}_{-14}$ is the angular velocity in units of $10^{-14}$ s$^{-1}$, and $n_4$ is the gas
density in units of $10^4$ \cmm\ (Menten \etal\ 1984).

Another parameter which measures the cloud stability is the ratio between the rotational energy and
combined thermal and non-thermal (turbulent) virial terms (e.g., Phillips 1999):
\begin{equation}
p' = 7.04\times10^{-2}\ R^2\ {\dot{\phi}}^2_{-14}\ {\Delta v}^{-2}\ ,
\label{Eq2b}
\end{equation}
where $R$ is the cloud radius in pc and $\Delta v$ the linewidth in \kms.
The influence of rotation and turbulence are comparable when
the stability parameter $p'(r) \approx 1$.

The calculations show that at the present stage of evolution of \object{L1251C}
($i$) rotational energy is a negligible fraction of the gravitational energy, $p \approx 0.02$, and
($ii$) turbulence exceeds the contribution due to rotation in determining cloud stability, $p' \approx 0.4$.

We now compare the \nhhh\ map with maps of C$^{18}$O and \nnhp.
As noted in Sec.~\ref{sect-1}, CO
molecules are usually distributed in larger volumes than \nhhh.
This is exactly what we observe in the case of \object{L1251C}.
Figure~\ref{fg6} shows that the ammonia core has a linear size, which is at least four times smaller,
$L_{\scriptscriptstyle {\rm NH}_3} \approx 0.2$ pc versus
$L_{\scriptscriptstyle {\rm C}^{18}{\rm O}} \approx 0.8$ pc.
The gas kinematics traced by ammonia differ noticeably from those revealed from C$^{18}$O: 
the observed linewidth-size relation 
\Dv$_{\scriptscriptstyle {\rm C}^{18}{\rm O}}$/\Dv$_{\scriptscriptstyle {\rm NH}_3} 
\approx 
(L_{\scriptscriptstyle {\rm C}^{18}{\rm O}} / L_{\scriptscriptstyle {\rm NH}_3})^{1.3}$
deviates significantly from the Larson law, \Dv\ $\propto L^{0.38}$ (Larson 1981)
established from $^{12}$CO(1--0) observations in molecular clouds over a wide range of scales 
($\sim 0.05-20$ pc).
This may imply that the core is not in perfect virial equilibrium and its
dynamical state is affected by the ambient environment as found in
numerical simulations of turbulent clouds (e.g., Kritsuk \etal\ 2013).

It is also noteworthy that the \nnhp(1--0) emission is distrubuted in a smaller area
than that of \nhhh(1,1), whereas the peak positions coincide. 
A direct comparison of the \nnhp\ map of Caselli \etal\ (2002)
with the ammonia map is shown in Fig.~\ref{fg6}{\bf c}.
(Both maps were obtained with similar spatial resolutions).
In general, \nhhh\ and \nnhp\ molecules trace the same volume elements in 
molecular cores (e.g. Tafalla \etal\ 2004), because they have a common chemical origin.
Nevertheless, our example hints at some chemical
differentiation in the process of the dynamical evolution of \object{L1215C}.

The mass of the central region, outlined by the thick contour in Fig.~\ref{fg6}{\bf c},
can be estimated using the same procedure as in Sect.~\ref{sect-3-1}. 
With the weighted mean linewidth $\Delta v = 0.31$ \kms, the mean gas density \nHH\ = $3.6\times10^4$ \cmm,
and the radius $R \sim 0.1$ pc, we have $M = 10.5M_\odot$ and $M_{\rm vir} = 2.3M_\odot$.
As in the case of \object{SS1},
the observed difference in the masses is probably caused by deviations from a uniform gas density
distribution and a core ellipticity.

\section{Autocorrelation and structure functions}
\label{sect-4}

\subsection{General considerations}
\label{sect-4-1}

In this Section we focus on the character of the observed stochastic motions of the gas
mapped in ammonia emission. Taking into account a low fractional abundance of \nhhh\
($X \la 10^{-7}$), the question arises whether
the ammonia emission traces the gas distribution. 
In what follows we assume that the answer is positive. To support this assumption,
we note that ammonia and submillimeter maps of dense cores show
a close correlation between the large-scale distributions of the \nhhh(1,1)
integrated intensity and the continuum emission as was pointed out in Sect.~\ref{sect-3} and
discussed in, e.g., Friesen \etal\  (2009).
Since dense cores are mostly composed
of molecular gas ($\sim 77$\% of H$_2$, and $\sim 23$\% of He by mass, neglecting metals),
we may consider ammonia  as a passive heavy impurity which does not perturb dynamics of the gas motion.
We suppose that the ammonia impurity is conservative, i.e., the relative abundance of \nhhh\ does not
change when moving from one point to another within the core.

Stochastic motion can be scale-free (purely random) or characterized by some length scale(s). In particular,
turbulence is a stochastic motion with multiple scales. An essential feature of turbulence is swirling 
motion which is viewed as eddies of various sizes.
The size of largest eddies is comparable with the physical dimension of the flow (in our case~---
of the molecular cloud itself), whereas a lower limit on the size of
eddies is of the order of the molecular mean free path,
i.e., for a dense core it is $\lambda = (\sqrt{2}n\sigma)^{-1} \sim 10^{-7}$ pc
(the effective cross section $\sigma \sim 10^{-15}$ cm$^2$, the gas density $n \sim 10^4$ \cmm). 
Actually, the size of the smallest eddies is determined by
viscosity which prevents the formation of eddies at lower scales.
At this limit the turbulent motion is converted into thermal
intermolecular energy which is then dissipated into the ambient medium.
The current paradigm regards turbulent flows as an energy cascade where the kinetic energy
is injected at large scales and is then transferred to smaller scales by
various nonlinear processes.
The model of energy cascade for the simplest case of the incompressible turbulent flow was first developed by
Kolmogorov (1941).
In reality the energy transfer is a rather complicated process. It is known to be highly intermittent 
at small scales due to the process of stretching of vortex filaments (e.g., Lesieur 1997).
Another important property is the multifractality of the  kinetic energy 
dissipation field characterized by
many scaling indices (e.g., Sreenivasan 1991).

To describe quantitatively the multiscale properties of the stochastic motions, 
different methods are used including calculations of
the two-point autocorrelation function (ACF) of the random velocity field,
the structure function (SF), and the power spectrum (PS)~--- the frequency-space
analog of the ACF (e.g., Kaplan \& Klimishin 1964; Scalo 1984; 
Dickman \& Kleiner 1985; Miesch \& Bally 1994; Lagrois \& Joncas 2011).

In what follows, we assume that the turbulent velocity field is homogeneous, 
i.e., the mean LSR radial velocity is constant within the \nhhh\ area,
and isotropic, 
i.e., ACF and SF depend only on the distance between two positions 
on the plane of the sky, $(\tilde{r} = |{\bf r}_1 - {\bf r}_2|)$.
In this case the normalized autocorrelation and structure functions are given by
(e.g., Yaglom 1987):
\begin{equation}
B(\tilde{r}) = \frac{\sum [V_c(r) - \mu][V_c(r + \tilde{r}) - \mu]}{\sigma^2_c\, N(\tilde{r})},
\label{Eq2}
\end{equation}
and
\begin{equation}
D(\tilde{r}) = \frac{\sum [V_c(r) - V_c(r + \tilde{r})]^2}{\sigma^2_c\, N(\tilde{r})}.
\label{Eq3}
\end{equation}
Here, $N(\tilde{r})$ is the number of pairs at each lag $\tilde{r}$.
The summation in (\ref{Eq2}) and (\ref{Eq3}) is over $N(\tilde{r})$ pairs. 
$V_c\ (\equiv V_{\scriptscriptstyle \rm LSR})$, defined in Eq.~(\ref{Eq3a}), 
is the mean radial velocity of the emitting gas averaged over the line-of-sight and given by the
velocity centroid of the \nhhh(1,1) line. 
$\mu$ is the mean centroid velocity:
\begin{equation}
\mu = \frac{\sum V_c({r})}{ N},
\label{Eq4}
\end{equation}
and $\sigma^2_c$ is the variance of centroid velocity fluctuations:
\begin{equation}
\sigma^2_c = \frac{\sum [V_c({r}) - \mu]^2}{ N} .
\label{Eq5}
\end{equation}
The total number of the velocity centroids in the map is $N$. 

For homogeneous fields, $B(\tilde{r})$ and $D(\tilde{r})$ are related as (Yaglom 1987): 
\begin{equation}
D(\tilde{r}) = 2[1 - B(\tilde{r})]\ .
\label{Eq7}
\end{equation}

From Eq.~(\ref{Eq2}) it follows that $B(\tilde{r}) \to 1$ at $\tilde{r} \to 0$.
With increasing $\tilde{r}$, the velocities become uncorrelated and 
$B(\tilde{r}) \to 0$ at $\tilde{r} \to \infty$. Correspondingly,
$D(\tilde{r}) \to 2$ at $\tilde{r} \to \infty$, and $D(\tilde{r}) \to 0$ at $\tilde{r} \to 0$. 
Sometimes, Eq.~(\ref{Eq7}) is used to justify homogeneity of the turbulent field
(e.g., Kitamura \etal\ 1993), i.e., if the calculated values of
ACF and SF are related in accord with (\ref{Eq7}), then the field is considered homogeneous.
However, mathematically it is not proven and, hence, the statement is ambiguous.   

As mentioned above, the ACF and PS provide the same information on the stochastic properties of the turbulent flow. 
For isotropic turbulence, 
the power spectrum $P(k)$ is a function of the wavenumber $k = 1/\tilde{r}$: 
\begin{equation}
P(k) = \frac{1}{2\pi} \int^\infty_{-\infty} {\rm e}^{-ik\tilde{r}} B(\tilde{r}) d\tilde{r} .
\label{Eq07}
\end{equation}
In the inertial range of the developed turbulence,
i.e., between the energy input and dissipation scales,
the PS may be approximated by a single power law with a spectral index $\kappa$, $P(k) \propto k^{-\kappa}$.  
Often observed for the cold interstellar medium are
3D spectra with $\kappa_{\scriptscriptstyle \rm 3D} \approx 3.6$ 
(note that for angle-averaged 1D spectra
$\kappa_{\scriptscriptstyle \rm 1D} = \kappa_{\scriptscriptstyle \rm 3D} - 2$), 
which is consistent with the Kolmogorov's scaling of incompressible turbulence, 
$P_{\scriptscriptstyle \rm 3D}(k) \propto k^{-11/3}$
(e.g., Falgarone \etal\ 2007).
In this case, the ACF is a monotonic function asymptotically approaching zero at large lags.

In practice, the calculated value of the ACF, $\zeta = B(\tilde{r})$, is affected by statistical errors.
We estimate the $1\sigma$ confidence interval for $\zeta$ with aid of the Fisher transformation:
\begin{equation}
f(z) = \frac{1}{2} \ln \frac{1 + z}{1 - z} = {\rm arctanh}(z) .
\label{Eq8}
\end{equation}
The function $f(z)$ follows 
approximately a normal distribution with the mean $m = f(\zeta) = {\rm arctanh}(\zeta)$ and
the standard error $\sigma_z = 1/\sqrt{N-3}$, where $N$ is the sample size.
Calculated in the transformed scale,
the $1\sigma$ confidence interval 
$(f_1 = m - \sigma_z, f_2 = m + \sigma_z)$ was converted back to the correlation scale 
\begin{equation}
z(f) = \frac{{\rm e}^{2f} -1}{{\rm e}^{2f} +1} = {\rm tanh(arctanh}(z))\ , 
\label{Eq9}
\end{equation}
yielding the $1\sigma$ confidence interval for each $\zeta$.

\subsection{ACFs for molecular cores}
\label{sect-4-2}

The calculated values of ACF and SF are not statistically reliable at large lags, where the
sample sizes are small, mainly due to
irregular boundaries of the core and the lack of measurements outside of it.
In our observations, 
the angular sizes of the \nhhh\ areas are not sufficiently large ($\la 10' \times 10'$)
compared with the angular resolution (FWHP = $40''$).
Therefore, to improve statistics we averaged the ACFs and SFs over azimuthal angle assuming isotropy,
as was usually done in previous studies
(e.g., Scalo 1984; Kitamura \etal\ 1993; Miesch \& Bally 1994),  
and calculated one-dimensional functions $B(\tilde{r})$ and $D(\tilde{r})$. 
Following Dickman \& Kleiner (1985) and Miesch \& Bally (1994),
these functions were corrected at nonzero lags for instrumental noise
giving rise to a velocity centroid variance $\sigma_{\rm rms}$. 
However, a typical value of $\sigma_{\rm rms} \sim 0.02$ \kms\ and 
the standard deviation of centroid velocity fluctuations $\sigma_c \sim 0.2$ \kms\
show that the correction factor $1 - (\sigma_{\rm rms} / \sigma_c)^2 \sim 1$ and thus
the influence of the instrumental noise on the autocorrelation function is negligible.

\begin{figure*}[t]
\vspace{0.0cm}
\hspace{-1.0cm}\psfig{figure=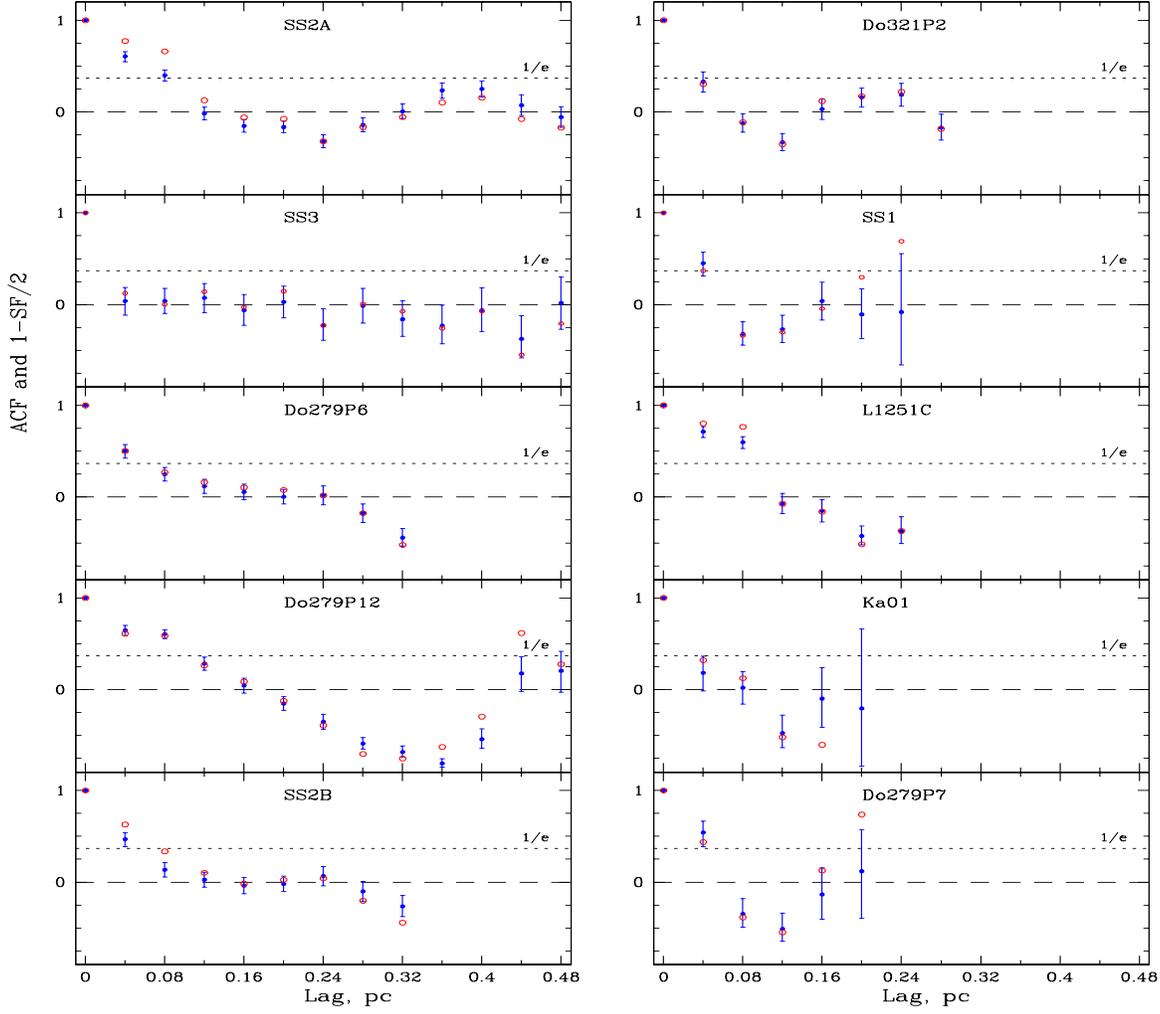,height=14.0cm,width=18.0cm}
\vspace{0.0cm}
\caption[]{Azimuthally averaged autocorrelation functions (ACFs)
plotted by blue dots with $1\sigma$ error bars
and azimuthally averaged structure functions (SFs)
plotted by red open circles for the \nhhh\ turbulent velocity fields. 
For the SFs, the values (1 -- SF/2) are shown to check homogeneity of the velocity map
in accord with Eq.~(\ref{Eq7}). 
The error bars for the (1 -- SF/2) points are the same as for the ACFs.
The zero and 1/e levels are shown by long- and short-dashed lines.
The sources \object{Do279P6}, \object{Do279P7}, \object{Do279P12}, \object{Do321P2}, \object{Ka01}, and
\object{SS3} are from Paper~I, while \object{SS1}, \object{SS2A}, \object{SS2B}, and \object{L1251C}~--- 
from the present paper. 
}
\label{fg7}
\end{figure*}

The profiles of the azimuthally averaged autocorrelation function $B(\tilde{r})$ and structure function
$D(\tilde{r})$ are shown in Fig.~\ref{fg7} where the latter is plotted in the form $1 - D(\tilde{r})/2$
for the comparison with the former. 
The cores were selected from both the current study and Paper~I based on
velocity maps having a sufficiently large number of observed positions to decrease
the statistical noise.
The scatter of the points plotted in Fig.~\ref{fg7} is mainly 
due to the statistical errors, 
it decreases with increasing numbers of data points. 
We simply cut the ACF profiles at large lags if
the scatter becomes too large, $|B(\tilde{r})| \ga 1$.  
Figure~\ref{fg7} shows a formal concordance 
between the ACF and SF values, so that
to analyze turbulence in molecular cores we mostly use the ACFs. 

The autocorrelation function gives us information on the characteristic scales of the turbulent
energy spectrum.
In particular, the correlation length, $\ell_c$, is a spatial scale corresponding to the lag where the ACF 
falls down to 1/e, $B(\ell_c) = {\rm e}^{-1}$.
If the turbulent structure of a core is completely resolved in observations, i.e., 
the correlation length is larger than
the beamwidth, then the true correlation length of the velocity 
field can be estimated.
However, in the case of partially resolved or unresolved observations, 
spurious correlations may occur
in the velocity field due to beam smoothing.
The relation between the apparent velocity correlation length, $\ell_a$, the true correlation length, $\ell_c$,
and the Gaussian beam width, $\theta$, of a radio telescope is given by (Kitamura \etal\ 1993):
\begin{equation}
\ell^2_a = \ell^2_c + \theta^2/\ln 4\ , 
\label{Eq10}
\end{equation}
which predicts for our ammonia observations $\ell_a \approx 34''$ ($\sim 0.02$ pc)
induced solely by the beam smoothing if the velocity scale of the underlying
motion is unresolved, $\ell_c \ll \theta$.

The scale $\ell_0$, where the correlation vanishes, $B(\ell_0) = 0$,
is another important parameter provided by ACFs. 
One may consider $\ell_0$ as
a typical size of the largest energetic turbulence fluctuations 
(e.g., Miville-Desch$\hat{\rm e}$nes \etal\ 1995).

The plots in Fig.~\ref{fg7} reveal three types of apparent ACF profiles: 
(1) spikelike at the origin as seen for the source \object{SS3};
(2) partially resolved with $\ell_a \sim \theta$
(\object{Do279P6}, \object{SS2B}, \object{Do321P2}, \object{SS1},
\object{Ka01}, \object{Do279P7}), and 
(3) resolved with $\ell_a > \theta$
(\object{SS2A}, \object{Do279P12}, \object{L1251C}).
Besides, seven ACFs have ranges with statistically significant negative correlations: \object{SS2A},
\object{Do279P12}, \object{Do321P2}, \object{SS1}, \object{L1251C}, \object{Ka01}, and \object{Do279P7}.

The spikelike ACF of \object{SS3} reflects a purely random motion  
with an uncorrelated velocity field (white noise process).
The core itself is located in the active star-forming region Serpens South,
harboring a number of young stellar objects (YSOs). 
It exhibits a complex velocity field structure and 
asymmetric ammonia lines (Fig.~10 and Fig.~B.8 in Paper~I).
However, an even higher star formation activity and asymmetric \nhhh\ lines are
observed towards, for example, \object{Do279P12} 
(Fig.~4 and Fig.~B.4 in Paper~I), where the presence of characteristic 
correlation scales is quite evident (Fig.~\ref{fg7}). 
Such a contrast between the velocity fields in similar objects,
with ${\cal M}_s \sim 1.5$ for both of them, is surprising
since it is assumed that turbulence is a universal property of most of the
molecular gas in the Milky Way from giant molecular clouds 
down in size to compact cores (McKee \& Ostriker 2007).

For partially resolved ACFs, we calculated the following apparent correlation lengths:
$\ell_a \approx 40''$ (\object{Do321P2}, \object{SS1}, \object{Ka01}),
$\ell_a \approx 50''$ (\object{Do279P7}), and
$\ell_a \approx 60''$ (\object{Do279P6}, \object{SS2B}).
The values corrected for beam smoothing are: $\ell_c \la 0.02$ pc, $\ell_c \sim 0.04$ pc, and 
$\ell_c \sim 0.05$ pc, respectively. 
The first zero crossing occurs at $\ell_0 \sim 0.06$ pc (\object{SS1}, \object{Do279P7}),
$\ell_0 \sim 0.08$ pc (\object{Do321P2}), $\ell_0 \sim 0.09$ pc (\object{Ka01}),
$\ell_0 \sim 0.16$ pc (\object{SS2B}), and $\ell_0 \sim 0.24$ pc (\object{Do279P6}).

All spatially resolved ACFs 
(\object{SS2A}, \object{L1251C}, \object{Do279P12})
show the correlation length
$\ell_c \sim 0.1$ pc, whereas $\ell_0$ is $\sim 0.13$, $\sim 0.15$, and $\sim 0.18$ pc, 
respectively.

\subsection{Oscillating ACFs}
\label{sect-4-3}

An interesting feature detected in about 70\% of ACFs 
of the \nhhh\ velocity maps is the presence of ranges with statistically significant negative correlation. 
This type of ACF deserves an additional discussion. 
We calculate ACFs based on maps obtained directly from \nhhh\ profile measurements
and under assumptions of spatial homogeneity and isotropy of the underlying velocity field.
However, it is known that slow drifts in the mean values of the stochastic field
and/or anisotropy can significantly distort the ACF, eventually producing regions of negative
correlations (Yaglom 1987). If we are interested in revealing the true statistical properties
of the stochastic field, then such bulk motions should be removed. In practice, this occurs by
filtering the measured maps with an appropriate window, the size of which is determined in
a series of consecutive trials (e.g, Lagrois \& Joncas 2011). 
Unfortunately, in our case this method is not applicable due to a low number of points in the maps,
so that the filtering with a large window becomes, in fact, equivalent to a simple
subtraction of the mean value of \VLSR. 
Nevertheless, there is a possibility to test the effects of bulk motions on the ACF
even with our limited maps:
namely, by comparing the ACFs calculated for two different directions.

\begin{figure}[t]
\vspace{0.0cm}
\hspace{-3.0cm}\psfig{figure=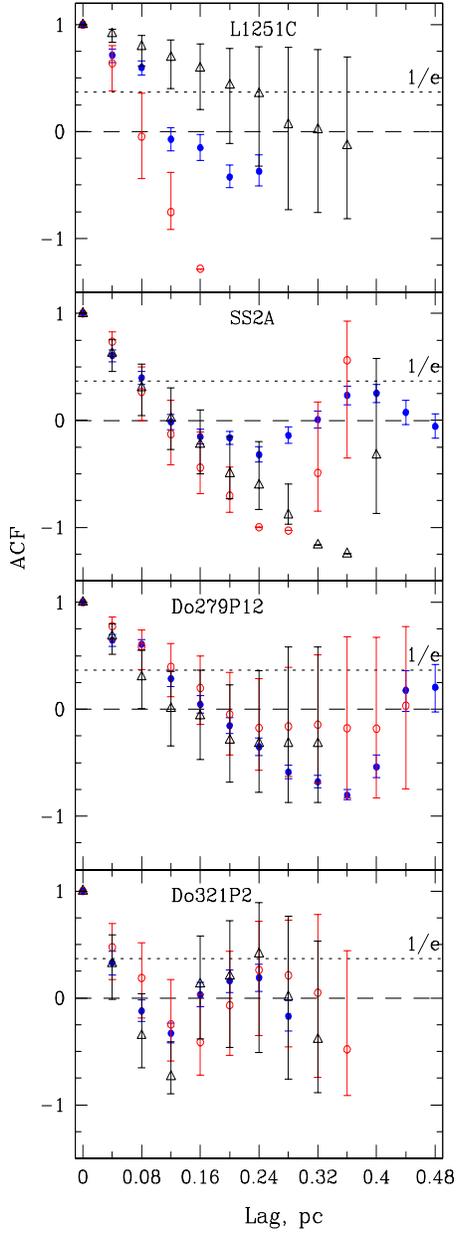,height=18.0cm,width=16.0cm}
\vspace{-1.0cm}
\caption[]{Testing the isotropy of the observed velocity field.
Blue dots show azimuthally averaged autocorrelation functions (ACFs)
same as already plotted in Fig.~\ref{fg7}, whereas 
ACFs calculated for directions $N \rightarrow S$ and $E \rightarrow W$
are shown by red open circles and black open triangles, respectively. 
Error bars show  $1\sigma$ statistical errors. 
At large lags, due to poor statistics, some points without error bars have a formal value
$| {\rm ACF} | \geq 1$ and therefore their errors cannot be defined using Eq.(\ref{Eq8}).  
In panel \object{L1251C},
a large-scale gradient of the velocity field (see Fig.~\ref{fg6}{\bf b}) 
manifests itself as a monotonically decreasing ACF with increasing lags
in the direction of the velocity gradient,
and as positive correlations for perpendicular direction. 
Both $N-S$ and $E-W$ ACFs deviate significantly from the azimuthally averaged ACF.
On the other hand, panels \object{SS2A}, \object{Do279P12}, and \object{Do321P2} show 
similar damping oscillations for all three ACFs
as expected if the velocity field is homogeneous and isotropic.
}
\label{fg8}
\end{figure}

In the present sample we have one object~--- \object{L1251C}~---
where the presence of the large-scale velocity gradient in the $N-S$ direction
across the whole surface area is clearly visible (Fig.~\ref{fg6}{\bf b}).
Figure~\ref{fg8} shows ACFs calculated along the velocity gradient, $B_{\scriptscriptstyle N-S}$,
and perpendicular to it, $B_{\scriptscriptstyle E-W}$.
$B_{\scriptscriptstyle N-S}$ decreases monotonically becoming negative at $\tilde{r} \ga 0.08$ pc,
whereas $B_{\scriptscriptstyle E-W}$ is everywhere positive.
On the other hand, cores \object{SS2A}, \object{Do279P12}, and \object{Do321P2}
do not demonstrate signs of large-scale motion. For them, both the 
the azimuthally averaged ACF and the
direction-based values $B_{\scriptscriptstyle N-S}$ and $B_{\scriptscriptstyle E-W}$
show similar behavior with oscillations damped with increasing lags (Fig.~\ref{fg8}).
Thus, we may conclude that for these three cores 
the assumption of isotropy is valid, and that the damping oscillations
indicate a real small-scale structure of the velocity field 
in the dense cores. 
  
Oscillating ACF are usually detected in atmospheric turbulence (Yaglom 1987);
in astrophysics, a similar shape of the azimuthally averaged ACF 
was reported for dense cores in the Milky Way by  Kitamura \etal\ (1993), where 
$4' \times 4'$ and $8' \times 8'$ areas towards the core \object{TMC-1C}
in Heiles' cloud 2 of the Taurus complex were mapped in the $^{13}$CO(1--0) and C$^{18}$O(1--0) lines.
Those observations were carried out
at the Nobeyama 45-m telescope with an angular resolution FWHP $ \approx 17''$ (0.01 pc).
As for extragalactic objects, oscillating ACFs were observed in two giant \ion{H}{ii} regions
\object{NGC~604} and \object{NGC~595} in \object{M33} by Medina Tanco \etal\ (1997) and 
Lagrois \& Joncas (2011), respectively.

In fluid dynamics, a similar behavior of the velocity autocorrelation function is 
predicted for a so-called Lennard-Jones fluid 
in which the interaction between a pair of flow particles is governed by the interplay of
the short-range repulsive and the long-range attractive forces
(e.g., Wijeyesekera \& Kushick 1979; Lad \& Pratap 2004).
Forces acting in molecular cores may be considered in a similar manner: 
gravity  generating the infall motion is an attractive force, whereas 
forces that provide support against gravitational collapse~--- centrifugal, dissipative and magnetic~---
are repulsive.

To describe the shape of the fluctuating  ACF we employ a model of damping oscillation
which is often used to represent 
empirical correlation functions which alternate between positive and negative values (Yaglom 1987):  
\begin{equation}
\tilde{B}(\tilde{r}) = {\rm e}^{-\gamma \tilde{r}}[\cos(a\tilde{r}) + \xi \sin(a\tilde{r})] .
\label{Eq11}
\end{equation}
The choice of this model, which is by no means unique, is determined by its simplicity and by
the fact that the gravitational potential within a spherically symmetric mass distribution 
behaves like $ \propto r^2$. 
Just for this type of the potential the ACF in the form of (\ref{Eq11}) 
was deduced for, e.g., a Brownian particle diffusion in a harmonic well (Glass \& Rice 1968). 
The parameters $\gamma$ and $a$ in (\ref{Eq11}) are defined in the spatial domain and, thus,
are measured in  pc$^{-1}$, the parameter $\xi$ is a dimensionless quantity.
The autocorrelation function $\tilde{B}(\tilde{r})$ is normalized to unity at lag $\tilde{r} = 0$. 

If the curve defined by Eq.~(\ref{Eq11}) closely fits experimental data, 
we can replace the empirical ACF
by an analytical function $\tilde{B}(\tilde{r})$ 
and, hence, to determine
the power spectrum also analytically. 
The Fourier transform, Eq.~(\ref{Eq07}), of $\tilde{B}(\tilde{r})$ gives 
\begin{equation}
P_{\scriptscriptstyle \rm 1D}(k) = 
\frac{1}{\pi}\frac{c_1 + c_2 k^2}{c_3 + c_4 k^2 + k^4} ,
\label{Eq16}
\end{equation}
where $c_1 = \gamma^3 + \gamma a^2 + \xi a \gamma^2 + \xi a^3$,
$c_2 = \gamma - \xi a$, 
$c_3 = (a^2 + \gamma^2)^2$,
$c_4 = 2(\gamma^2 - a^2)$,
and $P_{\scriptscriptstyle \rm 1D}(k) \propto k^{-2}$ at large $k$.

The Kolmogorov model of the velocity field in incompressible turbulence
postulates a dissipationless cascade characterized in the inertial range
by a transfer rate of kinetic energy independent of scale. In this model,
the power spectrum of the velocity $P_{\scriptscriptstyle \rm 1D} \propto k^{-5/3}$.
On the other hand, the spectrum of a highly compressible medium should scale
at high Mach numbers as $k^{-2}$ (e.g., Passot \etal\ 1988), which was confirmed in
numerical simulations by Kritsuk \etal\ (2007). 
One sees that the function defined by Eq.~(\ref{Eq16}) has the same asymptotic behavior at
large wavenumbers.

Below we consider the velocity field characteristics of three cores 
with oscillating ACFs in more detail 
employing the model (\ref{Eq11}) to fit their empirical 
ACFs in order to estimate the model parameters $\gamma$, $a$, and $\xi$.

\subsubsection{\object{SS2A} }
\label{sect-4-3-1}

The Aquila target \object{SS2A} exhibits a highly perturbed velocity field (Sect.~\ref{sect-3-2-1})
along with gas density variations (Table~\ref{tbl-1a}). It contains an IR source and
a few protostars which generate gas outflows seen from the enhanced linewidths of \nhhh\ 
in the vicinity of these sources.
No clear features of a rigid-body rotation of the core as a whole or any 
significant large-scale gradients of the velocity field
are found. 

\begin{figure}[t]
\vspace{0.0cm}
\hspace{-3.3cm}\psfig{figure=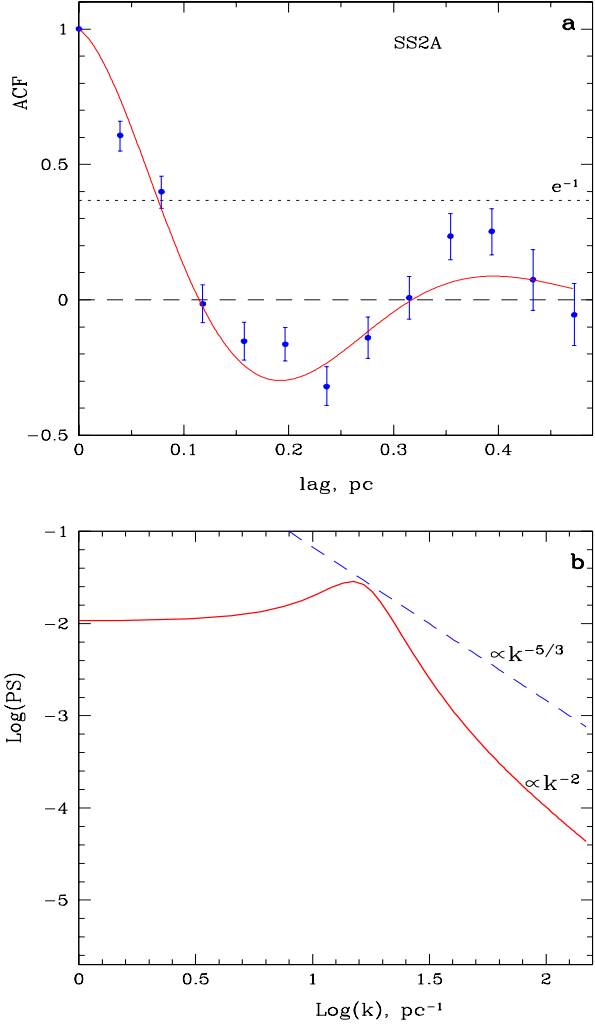,height=14.0cm,width=16.0cm}
\vspace{0.0cm}
\caption[]{The autocorrelation function (ACF) 
and the corresponding power spectra (PS) of the turbulence in \object{SS2A}.
({\bf a}) The solid line indicates the best fit of the  azimuthally averaged ACF 
(filled circles with $1\sigma$ error bars) to a model of decaying
oscillations discussed in Sect.~\ref{sect-4-3}.
The zero and e$^{-1}$ levels are shown by long- and short-dashed lines.
The lag $\tilde{r}$ is given in pc for the distance to the source $D = 203$ pc.
({\bf b}) The power spectrum of velocity $P_{\scriptscriptstyle \rm 1D}(k)$ 
(solid line) assuming
homogeneous and isotropic turbulence (the wavenumber $k = 1/\tilde{r}$). 
The spectrum is calculated using Eq.~(\ref{Eq16}).
Note the turnover of the PS at a scale of $\approx 0.06$ pc ($k_m \approx 17$ pc$^{-1}$),
where energy is being injected into the system.
For comparison,
the theoretical PS of a pure Kolmogorov cascade of incompressible turbulence 
is shown by a dashed line. The Kolmogorov
power-law index is $\kappa_{\scriptscriptstyle \rm 1D} = -5/3$, 
while the model value is $-2$ in the inertial range
at high wavenumbers, $k > k_m$. 
}
\label{fg9}
\end{figure}

Figure~\ref{fg9}{\bf a} shows the best fit of the model curve (solid line) to  
the measured velocity correlation function (filled circles with $1\sigma$ error bars). 
In panel {\bf b}, the corresponding power spectrum given by equation (\ref{Eq16})
with the estimated model parameters $\gamma \simeq 6$ pc$^{-1}$, $\xi \simeq 0.2$ and
$a \simeq  15$ pc$^{-1}$ is plotted.
For comparison, a pure Kolmogorov cascade of incompressible turbulence (dashed line) with a scaling 
$\propto k^{-5/3}$ is also presented. 
The power spectrum of \object{SS2A}
forms a noticeable knee at $k_m \approx 17$ pc$^{-1}$.
In general, three ranges in the model PS shape can be distinguished:
the environmental ($k < k_m$), driving  ($k \approx k_m$), and inertial/dissipative 
PS ($k \ga k_m$)\footnote{We note 
that the meaning of the inertial range is applicable to a well developed turbulence which is characterized 
by statistical regularities over a wide range of scales. This
may not be the case, however, for a compact dark core subjected to self-gravitation.}.
In the environmental range (at large scales), 
the model shows a flat spectrum at different wavenumbers. 
The energy is injected into the gas motion at $k \sim k_m$ ($\ell_m \sim 0.06$ pc)  
and is dissipated at scales $k > k_m$.
The model ACF drops to the level of $1/{\rm e}$ at $\ell_c = 0.08$ pc (Fig.~\ref{fg9}{\bf a}).

\subsubsection{\object{Do279P12} }
\label{sect-4-3-2}

The source \object{Do279P12} is a dark core with numerous embedded YSOs,
IR and Herbig-Haro objects along with submm-continuum and radio sources (Fig.~4 in Paper~I).  
The ammonia lines show a double component structure towards a few offsets
(Figs.~6 and B.4 in Paper~I). We interpreted this splitting 
as a purely kinematic effect since it is present in the optically thin \nhhh\ lines as well.
Both core \object{SS2A} and \object{Do279P12} have similar linear sizes and morphology
of the \nhhh(1,1) map: an elongated surface of irregular form.

\begin{figure}[t]
\vspace{0.0cm}
\hspace{-3.3cm}\psfig{figure=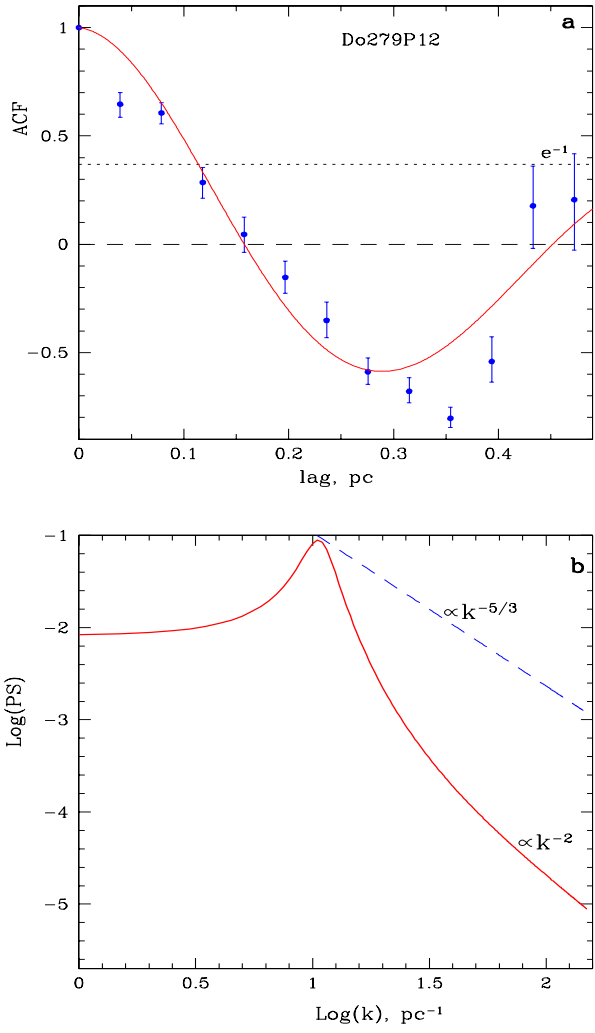,height=14.0cm,width=16.0cm}
\vspace{0.0cm}
\caption[]{Same as Fig.~\ref{fg9} but for the source Do279P12.
}
\label{fg10}
\end{figure}

The correlation length $\ell_c \approx 0.12$ pc (Fig.~\ref{fg10}{\bf a})
is well defined,
thus suggesting a highly structured velocity field.
The ACF shows a decaying oscillation in the range (0.16, 0.5) pc
with a prominent negative region between 0.16 pc and  0.44 pc, 
while the corresponding PS (Fig.~\ref{fg10}{\bf b})
exhibits a sharp peak at $k_m \approx 11$ pc$^{-1}$
($\ell_m \approx 0.09$ pc) which is of the order of the sonic length, $\ell_s \sim 0.1$ pc.
The estimated model parameters  
are $\gamma \simeq 1.8$ pc$^{-1}$, $a \simeq 11$ pc$^{-1}$,
and $\xi \simeq 0.1$.

\subsubsection{\object{Do321P2} }
\label{sect-4-3-3}

The ACF profile of \object{Do321P2} and those
shown in the right panels of Fig.~\ref{fg7}  (\object{SS1}, \object{Ka01}, \object{Do279P7}) 
are only partially resolved and affected at small scales by beam smoothing.
However, at larger scales,
their ACFs demonstrate statistically significant negative correlations resembling those
measured in \object{SS2A} and \object{Do279P12}.
\object{Do321P2} has a larger angular size and, thus, its ACF was measured more accurately. 

\object{Do321P2} is a filamentary dark core with a few YSOs, SCUBA, and IRAS
sources embedded (Fig.~7 in Paper~I).
The \nhhh\ velocity map (Fig.~7{\bf b} in Paper~I) shows an irregular structure
with distinct pockets of redshifted and blueshifted ammonia
emission around the clusters of YSOs. The observed kinematics are caused by
large-scale motions in this star-forming region. 

\begin{figure}[t]
\vspace{0.0cm}
\hspace{-3.3cm}\psfig{figure=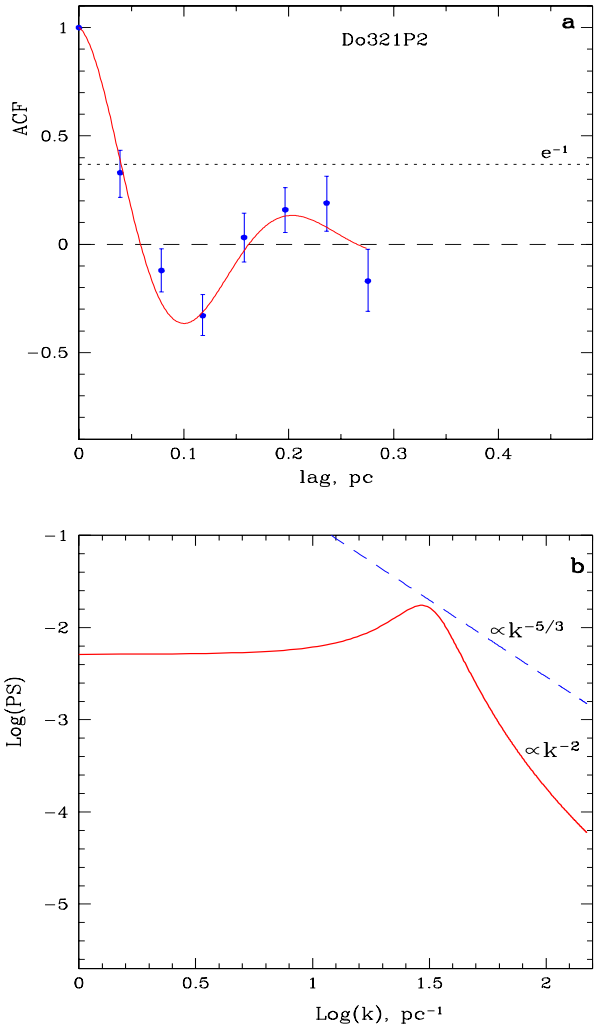,height=14.0cm,width=16.0cm}
\vspace{0.0cm}
\caption[]{Same as Fig.~\ref{fg9} but for the source Do321P2.
}
\label{fg11}
\end{figure}

The azimuthally averaged ACF of the velocity
field is shown in Fig.~\ref{fg11}{\bf a} by filled circles with $1\sigma$ error bars. The smooth line
represents the model ACF.  The estimated model parameters are
$\gamma \simeq 9.8$ pc$^{-1}$, $a \simeq 30.4$ pc$^{-1}$, and $\xi \simeq 0.2$. 
In panel {\bf b}, the power spectrum is shown. 
The model results may be affected by beam effects  
and caution must be taken in prescribing them any significance. 
Higher angular resolution observations are required to get reliable physical parameters of the
bulk motion in this and in the above mentioned cores
showing similar properties  (\object{SS1}, \object{Ka01}, and \object{Do279P7}).

\section{Summary}
\label{sect-5}

We have used the Effelsberg 100-m telescope to map 
\nhhh(1,1) and (2,2) toward three dark cores in the 
Aquila rift cloud complex (\object{SS1}, \object{SS2}A, B) 
and toward one core in the Cepheus molecular cloud (\object{L1251C}).
The Aquila targets were preliminary selected from the CO,
$^{13}$CO, C$^{18}$O survey carried out with the Delingha 14-m telescope (Wang 2012).
The Cepheus target was delineated from the C$^{18}$O map by Sato \etal\ (1994).

The observed ammonia cores demonstrate complex intrinsic motions 
and fluctuating gas density and kinetic temperature on scales $0.04 \la \ell \la 0.5$ pc.
The amplitude of the gas density fluctuation is
$(n_{\scriptscriptstyle {\rm H}_2})_{\rm max}/(n_{\scriptscriptstyle {\rm H}_2})_{\rm min} = 2$, and 4 for,
respectively, \object{SS1} and \object{SS2A}, and it is up to
5 for both \object{SS2B} and \object{L1251C},
while the mean density in these core is 
$\langle n_{\scriptscriptstyle {\rm H}_2} \rangle/10^4 = 2.4$, 1.8, 1.7,
and 3.6 \cmm, respectively (Tables~\ref{tbl-1a}, \ref{tbl-2a}).
The measured ammonia abundances, $X=$ [\nhhh]/[H$_2$], 
vary between $X = 2\times10^{-8}$ (\object{SS1} and \object{L1251C})
and $X = 10^{-7}$ (\object{SS2}A, B). 

A systematically higher kinetic temperature is measured in cores harboring YSOs 
(\object{SS1}, \object{SS2A}, \object{L1251C}) as compared to the starless object \object{SS2B}: 
maximum \Tkin\ $\simeq 15$~K for the former versus maximum \Tkin\ $\simeq 11$~K for the latter.
The minimum \Tkin\ $= 8.8$~K is detected in \object{SS2B} as well.

The measured kinetic temperature was used to estimate the line of sight dispersion 
of the non-thermal bulk motions from the observed linewidths.
We found that the Cepheus core differs noticeably 
with respect to this parameter from the Aquila sources.
A quiet gas motion with $\sigma_{\rm turb} \sim v_{\rm th}$ is observed towards \object{L1251C},
while the velocity fields are highly perturbed in Aquila. The two Aquila cores \object{SS1} and
\object{SS2A} have internal regions where the Mach number ${\cal M}_s > 1$ and
the velocity dispersion changes with a high rate of
$d\sigma_{\rm turb}/dr \ga 4$ km~s$^{-1}$~pc$^{-1}$. The third core \object{SS2B} with
${\cal M}_s < 1$ has two regions where $d\sigma_{\rm turb}/dr \ga 2$ km~s$^{-1}$~pc$^{-1}$.
However, $d\sigma_{\rm turb}/dr \approx 0$ km~s$^{-1}$~pc$^{-1}$ for \object{L1251C}
where ${\cal M}_s < 1$ as well.

\object{L1251C} is also the only core among those analyzed in this paper where a regular velocity gradient 
with respect to the central `coherent' velocity zone is observed.
When interpreted as rigid-body rotation, this gives an angular velocity
$\dot{\phi} \approx 7\times10^{-14}$ s$^{-1}$. 
However, the corresponding rotational energy is a negligible fraction of the
gravitational energy since the parameter $p$ in Eq.~(\ref{Eq1b}) is $\ll 1$. 
In the same way,
a significant deviation from Larson's law ($\Delta v \propto L^{1.3}$ instead of $\propto L^{0.38}$)
indicates a decreasing contribution of the turbulent energy to core stability.
Another distinct feature of this core is the noncongruent maps of
\nhhh\ and N$_2$H$^+$  emission showing that these molecules do not trace
exactly the same region and that the latter is concentrated within a smaller volume surrounding
the central gas density peak \nhhh(1,1).

Masses from one to ten $M_\odot$ are estimated for 
\object{SS1} and \object{L1251C}, whereas they are $\la M_\odot$ 
for the $\alpha$ and $\beta$ peaks in
\object{SS2A} and \object{SS2B}, and
significantly less than the solar mass for other ammonia peaks in these cores. 

We present new examples of 
the autocorrelation function (ACF) of the velocity field
with slowly decaying oscillations on scales $\sim 0.04-0.5$ pc. 
The corresponding power spectrum (PS) deviates significantly from a single power-law cascade 
and exhibits a knee in its shape. Such a kind of the PS may arise at a transient stage of the dynamic 
evolution of a dense core when self-gravity directly competes with the nonconservative forces.
We suggest that oscillating ACFs may indicate a damping of the developed turbulent
flows surrounding the dense but less turbulent core~--- a transition to dominating gravitational
forces and, hence, to gravitational collapse.
 
To conclude, we would like to emphasize that 
while the damping oscillations in the autocorrelation function
may reflect a particular turbulent structure of the dense cores, 
their reliability requires further confirmations
from higher angular resolution observations and better statistics. 
The study of Aquila dense cores will be continued with
the Effelsberg 100-m telescope to investigate deeply their physical and kinematic properties.

\begin{acknowledgements}
We thank the staff of the Effelsberg 100-m telescope for their assistance in observations, and
we appreciate Alexei Kritusk's and Andrei Bykov's comments on an early version of the manuscript.
We also thank the anonymous referee for suggestions that led to substantial improvements of the paper.
SAL is grateful for the kind hospitality of 
the Max-Planck-Institut f{\"u}r Radioastronomie and Hamburger Sternwarte 
where this work has been prepared.  
This work was supported in part by the grant DFG
Sonderforschungsbereich SFB 676 Teilprojekt C4, and
by the RFBR grant No.~14-02-00241.
\end{acknowledgements}

\clearpage

\clearpage

\clearpage

\clearpage

\clearpage

\clearpage

\clearpage

\clearpage

\clearpage

\clearpage

\clearpage

\clearpage
\begin{table*}[t!]
\centering
\caption{Peak intensities of \nhhh(1,1) line emission towards \object{SS1}, \object{SS2}, and \object{L1251C}
}
\label{tbl-1}
\begin{tabular}{l c c c r@{,}l c r@{.}l r@{.}l l }
\hline
\hline
\noalign{\smallskip}
\multicolumn{1}{c}{Source} & Peak 
& \multicolumn{2}{c}{Position}& \multicolumn{2}{c}{Offset} & 
\multicolumn{1}{c}{\Tmb$^b$} & \multicolumn{2}{c}{$V_{\scriptscriptstyle\rm LSR}$} 
& \multicolumn{2}{c}{$\Delta v^c$} & 
\multicolumn{1}{c}{Date} \\[-2pt]
 & \multicolumn{1}{c}{Id.$^a$} & \multicolumn{1}{c}{$\alpha_{2000}$} & \multicolumn{1}{c}{$\delta_{2000}$} 
& {$\Delta\alpha$} & {$\Delta\delta$} 
& \multicolumn{1}{c}{(K)} & \multicolumn{2}{c}{(\kms)} & \multicolumn{2}{c}{(\kms)} 
& \multicolumn{1}{c}{(d-m-y)} \\
&  & ($^{\rm h}$: $^{\rm m}$: $^{\rm s}$) & ({\degr}: {\arcmin}: {\arcsec}) &
({\arcsec})&({\arcsec})  \\
\noalign{\smallskip}
\hline

\noalign{\smallskip}
\object{SS1} &    & 18:29:09.6 & $-$02:05:40 & $0$&$0$ & 0.9(1) & 6&41(1) & 0&69(2) & 16-03-13 \\ 
& $\alpha$       &             &            & $-40$&$0$& 1.1(1) & 6&139(6)& 0&54(1) & 16-03-13  \\ 
&        &             &            & $-40$&$-40$& 0.8(1) & 6&32(2)& 0&95(5) & 16-03-13  \\ 

\noalign{\smallskip}
\object{SS2A} &$\eta$ & 18:29:05.1 & $-$01:42:00 & $0$&$0$  & 0.8(1) & 7&05(1) & 1&04(3) &  18-03-13, 09-05-13  \\ 
& $\alpha$        &          &          & $-40$&$200$   & 2.9(3) & 7&361(6) & 0&47(1)& 22-03-13 \\ 
& $\beta$         &          &          & $-80$&$-40$   & 2.4(2) & 6&622(3) & 0&416(6)& 18-03-13, 22-03-13 \\ 
& $\gamma$        &          &          & $-360$&$320$   & 2.0(2) & 6&457(4) & 0&24(1)& 23-03-13 \\ 
& $\delta$        &          &          & $-240$&$240$   & 1.9(2) & 6&781(6) & 0&30(1)& 23-03-13 \\ 
& $\varepsilon$   &          &          & $-240$&$440$   & 1.4(1) & 7&175(7) & 0&22(2)& 23-03-13 \\ 
& $\zeta$   &          &                & $120$&$120$   & 1.1(1) & 7&526(9) & 0&24(2)& 23-03-13 \\ 
& $\theta$   &          &               & $-80$&$-160$  & 0.8(1) & 8&24(3) & 0&46(2)& 23-03-13 \\ 

\noalign{\smallskip}
\object{SS2B} &$\alpha$ & 18:29:47.8 & $-$01:37:20 & $0$&$0$  & 2.3(2) & 8&187(2) & 0&259(5) &  24-03-13, 09-05-13  \\ 
& $\beta$         &          &          & $40$&$40$   & 2.3(2) & 8&322(2) & 0&287(4)& 24-03-13, 09-05-13 \\ 
& $\gamma$        &          &          & $240$&$80$  & 2.2(2) & 8&178(3) & 0&426(6)& 09-05-13 \\ 
& $\delta$        &          &          & $120$&$0$   & 1.7(2) & 8&048(3) & 0&26(1)& 08-05-13 \\ 
& $\varepsilon$   &          &          & $-20$&$100$ & 1.6(2) & 8&14(1) & 0&28(2)& 24-03-13 \\ 

\noalign{\smallskip}
\object{SS2C} &$\alpha$ & 18:31:09.2 & $-$01:34:40 & $0$&$0$  & 0.4(1) & 8&49(3) & 0&25(3) &  09-05-13  \\ 

\noalign{\smallskip}
\object{L1251C} &$\alpha$ & 22:35:53.6 & $+$75:18:55 & $0$&$0$  & 3.9(2) & $-$4&722(3) & 0&290(7) &  
23/24-03-13, 09-05-13  \\ 
\noalign{\smallskip}
\hline
\noalign{\smallskip}
\multicolumn{12}{l}{{\bf Notes.} $^a$Greek letters mark peaks of ammonia emission indicated in Figs.~\ref{fg2}, \ref{fg4},
\ref{fg5}, and \ref{fg6}. } \\
\multicolumn{12}{l}{ $^b$The numbers in parentheses correspond to a $1\sigma$ statistical error 
on the last digit.  }\\
\multicolumn{12}{l}{ $^c$Linewidth (FWHP). }  
\end{tabular}
\end{table*}

\clearpage
\Online
\begin{appendix}
\section{Ammonia spectra towards \object{SS1}, \object{SS2}, and \object{L1251C}
and derived physical parameters}

The ammonia spectra were analyzed in the same way as in Paper~I.
The radial velocity $V_c\ (\equiv $\VLSR),  the linewidth \Dv\
(FWHP), the optical depths \ta\ and \tb, the integrated ammonia
emission $\int$\Tmb$dv$, and the kinetic temperature \Tkin, are well
determined physical parameters, whereas the excitation temperature
\Tex, the ammonia column density $N$(\nhhh), and the gas
density \nHH\ are less certain since they depend on the  beam filling factor $\eta$,
which is not known for unresolved cores (see Appendix~A in Paper~I).

The $1\sigma$ errors of the model parameters were estimated from the diagonal elements of the
covariance matrix calculated for the minimum of $\chi^2$. 
The error in $V_c$ was also estimated independently by the $\Delta \chi^2$ method (e.g.,
Press \etal\ 1992) to control both results. 
When the two estimates differed, the larger error was adopted.
An independent control gives an analytical estimate of the uncertainty of 
the Gaussian line center by Landman \etal\ (1982):
\begin{equation}
\sigma_{\scriptscriptstyle V} \approx 0.69 (rms/T_{\scriptscriptstyle \rm MB})
\sqrt{\Delta_{\rm ch}\cdot\Delta v}\, ,
\label{Eq1}
\end{equation}
where $\Delta_{\rm ch}$ and $\Delta v$ are the channel width and the linewidth, respectively,
and the parameter $rms$ is the root mean square noise level. 

The results of our analysis are presented in Tables~\ref{tbl-1}, \ref{tbl-1a}, and \ref{tbl-2a}.
In Tables~\ref{tbl-1a} and \ref{tbl-2a}, Cols.~11 and 12 list the total optical depth 
$\tau_{\rm tot}$ [see Eq.~(\ref{Eq3a})] which is the maximum optical depth that an unsplit $(J,K) = (1, 1)$ or (2, 2) 
rotational line would have at the central frequency if the hfs levels were populated with the same excitation
temperature for the two lines (1, 1) and (2, 2). 
Column 13 lists the total column density $N_{\rm tot}$(\nhhh) defined in Eq.~(\ref{Eq4a}).

For high signal-to-noise (S/N) data ($rms \sim 0.05$~K per 0.039 \kms\ channel 
and \Tmb\ $\sim 2$~K) and a narrow line (\Dv\ $\sim 1$ \kms),
Eq.~(\ref{Eq1}) gives $\sigma_{\scriptscriptstyle V} \sim 0.003$ \kms, which is in line with the precision
estimated from the covariance matrix.
However, the accuracy of the line position centering is probably a few times
lower taking into account irregular shifts ($\sim \frac{1}{4}\Delta_{\rm ch}$)
of the radial velocities $V_c$ measured in our Effelsberg spectra (Levshakov \etal\ 2013b).

The errors of \Dv\ depend on the S/N ratio and vary from $\sim 0.005$ \kms\ for strong ammonia lines
(S/N $\ga 30$) to $\sim 0.03$ \kms\ if S/N $\la 10$.
Since the uncertainty in the amplitude scale
calibration was $\sim20$\% (Sect.~\ref{sect-2}), the same order of magnitude errors are obtained 
for \Tmb, \Trot, \Tkin, \ta, and \tb. 
For \Tex, $N_{\rm tot}$(\nhhh), and \nHH, 
we estimated lower bounds corresponding to the filling factor $\eta = 1$ (Paper~I).

\begin{figure}
\vspace{0.0cm}
\hspace{-1.0cm}\psfig{figure=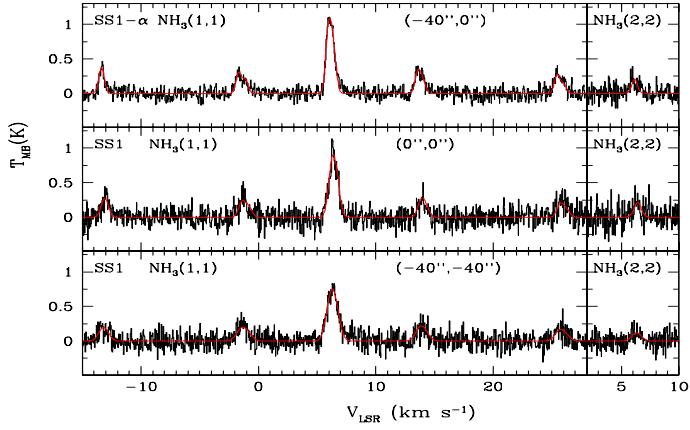,height=16.0cm,width=20.0cm}
\vspace{-10.0cm}
\caption[]{
Ammonia \nhhh(1,1) and (2,2) spectra 
towards the molecular core \object{SS1}.
The channel spacing is 0.039 \kms, the spectral resolution 
(Full Width at Half Peak, FWHP) is 0.045 \kms. 
The red curves show the fit of a one-component Gaussian model
to the original data.
}
\label{fg1a}
\end{figure}

\begin{figure}
\vspace{0.0cm}
\hspace{-1.0cm}\psfig{figure=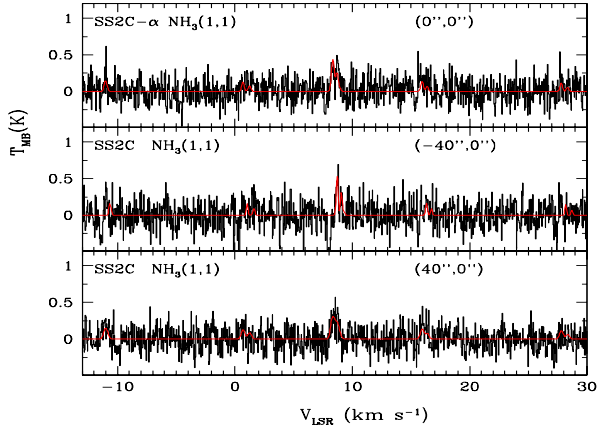,height=16.0cm,width=20.0cm}
\vspace{-10.0cm}
\caption[]{
Inverse ammonia \nhhh(1,1) spectra (i.e., $-1\times T_{\rm MB}$, where $T_{\rm MB}$
is the originally measured  main beam brightness temperature) 
towards the second
serendipitously detected molecular core \object{SS2C} to show
its signals, obtained from off-positions, in emission.
The ``absorption'' feature seen in the blue wing of the central \nhhh\ 
line at offsets ($0''$,$0''$), ($-40''$,$0''$), and ($40''$,$0''$)
is the Doppler shifted ammonia emission line detected at offsets
($320''$,$440''$), ($280''$,$440''$),  and ($360''$,$440''$) towards 
the source \object{SS2B}.
The channel spacing is 0.039 \kms, the spectral resolution FWHP = 0.045 \kms. 
The red curves show the fit of a one-component Gaussian model
to the original data. 
}
\label{fg2a}
\end{figure}

\begin{figure*}
\vspace{0.0cm}
\hspace{-1.0cm}\psfig{figure=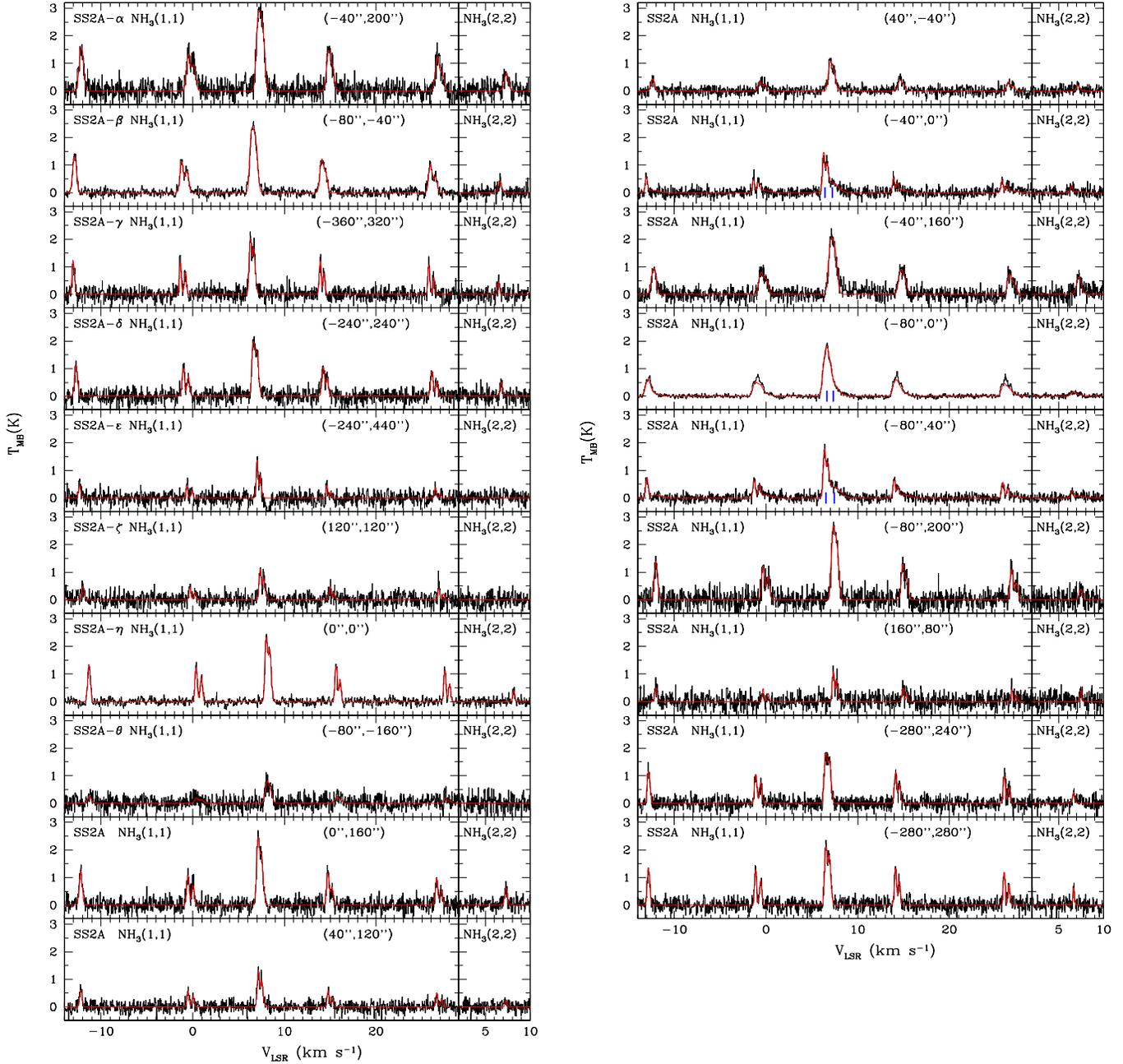,height=18.0cm,width=19.0cm}
\vspace{0.0cm}
\caption[]{
Ammonia \nhhh(1,1) and (2,2) spectra towards the molecular core \object{SS2A}.
The channel spacing is 0.039 \kms, the spectral resolution FWHP = 0.045 \kms. 
The red curves show the fit of a one- or two-component Gaussian model (the latter is marked by 
two vertical ticks at the position of the central \nhhh\ line)
to the original data.
}
\label{fg3a}
\end{figure*}

\begin{figure*}
\vspace{0.0cm}
\hspace{-1.0cm}\psfig{figure=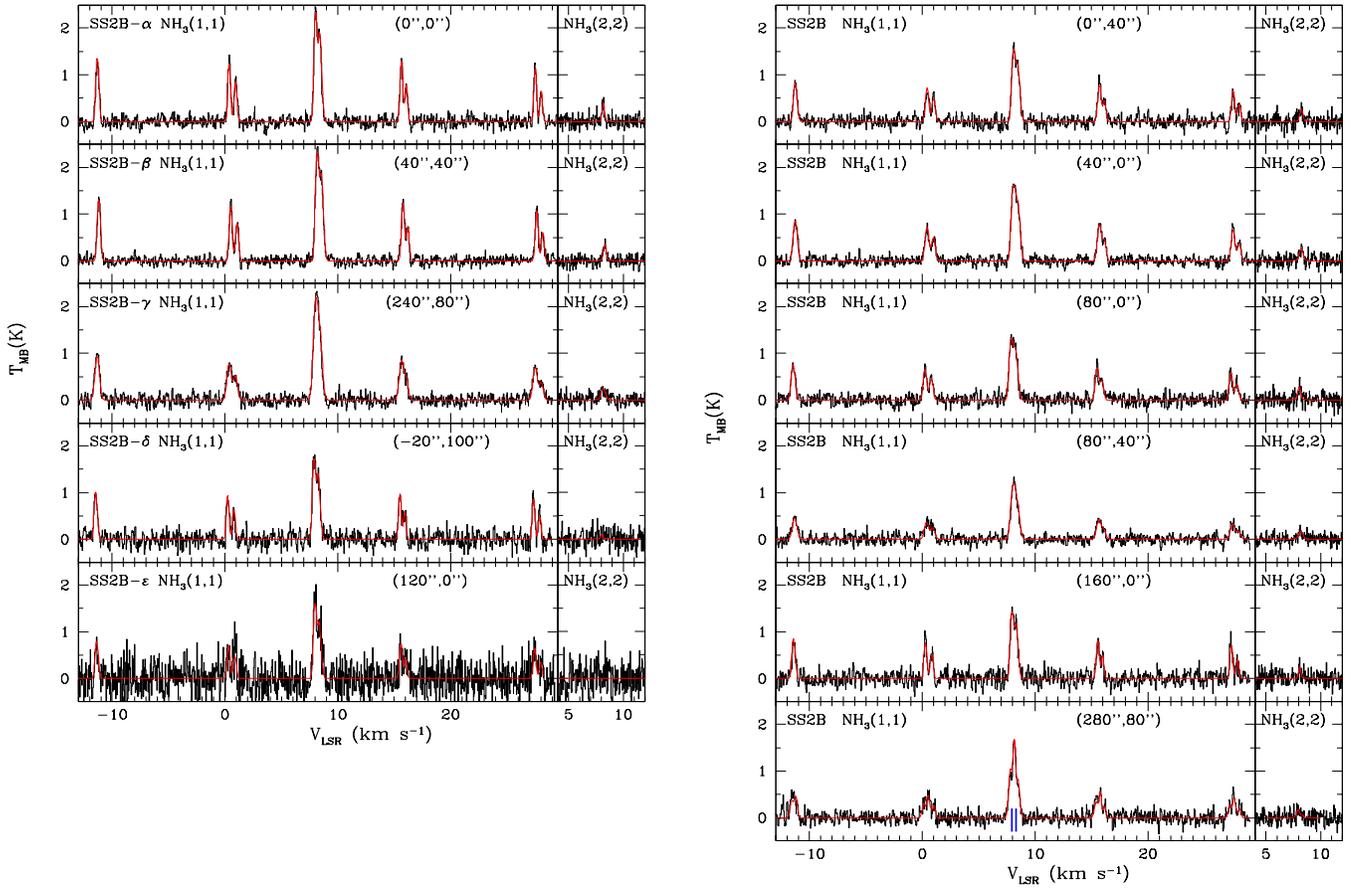,height=18.0cm,width=19.0cm}
\vspace{-5.0cm}
\caption[]{
Same as Fig.~\ref{fg3a} but for the serendipitously detected
starless molecular core \object{SS2B}.
}
\label{fg4a}
\end{figure*}

\begin{figure*}
\vspace{0.0cm}
\hspace{0.0cm}\psfig{figure=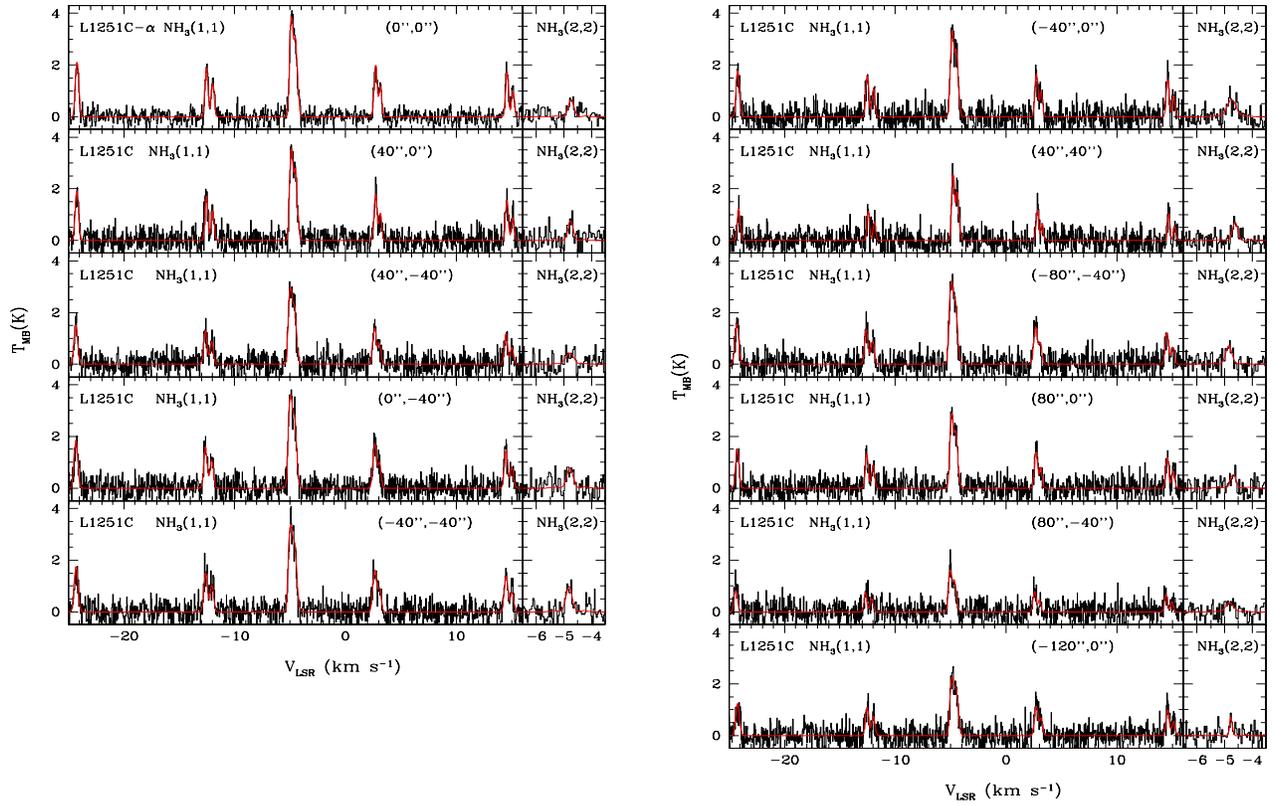,height=16.0cm,width=18.0cm}
\vspace{-4.5cm}
\caption[]{
Ammonia \nhhh(1,1) and (2,2) spectra towards the source \object{L1251C}.
The channel spacing is 0.039 \kms, the spectral resolution FWHP = 0.045 \kms. 
The red curves show the fit of a one-component Gaussian model 
to the original data. The model parameters are listed in Table~\ref{tbl-2a}.
}
\label{fg5a}
\end{figure*}

Examples of the \nhhh(1,1) and (2,2) spectra observed towards \object{SS1}, \object{SS2A},
\object{SS2B}, \object{SS2C}, and \object{L1251C} are shown in Figs.~\ref{fg1a}-\ref{fg5a}.
The measured parameters at the positions with both detected 
inversion transitions (1,1) and (2,2) are listed in Tables~\ref{tbl-1a} and \ref{tbl-2a}.

The physical parameters such as the total optical depth of an inversion line,
$\tau_{\rm tot}$, the radial velocity, $V_c$, the linewidth, \Dv,
and the amplitude, ${\cal A}$, were estimated by fitting a Gaussian model, $T_{\rm syn}(v)$, 
to the observed spectrum, $T_{\rm obs}(v)$, 
by means of a $\chi^2$-minimization procedure:
\begin{equation}
\chi^2 = \sum\, \left[ T_{\rm syn}(v) - T_{\rm obs}(v) \right]^2/{rms}^2\ ,
\label{Eq1a}
\end{equation}
where
\begin{equation}
T_{\rm syn}(v) = {\cal A} \{ 1 - \exp\left[-\tau(v) \right] \},
\label{Eq2a}
\end{equation}
and
the optical depth $\tau(v)$ at a given radial velocity $v$ is 
\begin{equation}
\tau(v) = \tau_{\rm tot} \sum^n_{i=1} r_i \exp \{
-2.773 [(v - V_c) + v_i]^2 / (\Delta v)^2 \}.
\label{Eq3a}
\end{equation}
Here $n$ is the number of magnetic hfs components of the inversion transition
[$n = 18$ and 21 for the $(J,K) = (1,1)$ and (2,2) inversion transitions],
$r_i$ is the relative intensity of the $i$th hfs line, and $v_i$ is its
velocity separation from the fiducial frequency (these parameters
are taken from Kukolich 1967, and Rydbeck \etal\ 1977).

The total \nhhh\ column density is defined by
\begin{eqnarray}
N({\rm NH}_3) & = & N_{11}\cdot [(1/3)\exp(23.2/T_{\rm rot}) + 1 + \nonumber\\
       &  &  + (5/3)\exp(-41.5/T_{\rm rot}) + \nonumber\\
       &  &  + (14/3)\exp(-105.2/T_{\rm rot}) + \ldots ] ,
\label{Eq4a}
\end{eqnarray}
where the relative population of all metastable levels of both ortho-\nhhh\
($K = 3$), which is not observable,
and para-\nhhh\   ($K = 1, 2$) is assumed to be
governed by the rotational temperature \Trot\ of the system at thermal equilibrium
(Winnewisser \etal\ 1979).

If both inversion lines \nhhh(1,1) and (2,2) are detected, we can estimate the kinetic 
temperature \Tkin\ [Eq.~(A.15) in Paper~I]
and find the thermal contribution to the observed linewidth, 
$v_{\rm obs} = $\Dv/$2\sqrt{\ln 2}$. 
The thermal velocity is defined by
\begin{equation}
v_{\rm th} = \sqrt{\frac{2k_{\scriptscriptstyle \rm B}T_{\rm kin}}{m}} ,
\label{Eq5a}
\end{equation}
where $k_{\scriptscriptstyle \rm B}$ is the Boltzmann constant, and $m$ is the mass of a particle.
For \nhhh, it is $v_{\rm th} = 0.03\sqrt{T_{\rm kin}}$ (\kms). 

If the contribution of the thermal $v_{\rm th}$ and non-thermal (turbulent) $v_{\rm turb}$
components to the linewidth are independent of one another, then
\begin{equation}
v_{\rm obs} = \sqrt{v^2_{\rm th} + v^2_{\rm turb}} ,
\label{Eq6a}
\end{equation}
and the non-thermal velocity dispersion $\sigma_{\rm turb}$ along the line of sight is
$\sigma_{\rm turb} = v_{\rm turb}/\sqrt{2}$.
This value can be compared with the thermal sound speed
\begin{equation}
c_s = (\tilde{\gamma} P_{\rm th}/\rho)^{1/2} ,
\label{Eq7a}
\end{equation}
where $P_{\rm th}$ is the thermal pressure and $\rho$ the gas density.
For an isothermal gas ($\tilde{\gamma} = 1$), $c_s = 0.06\sqrt{T_{\rm kin}}$ (\kms).  

The Mach number is defined locally as
\begin{equation}
{\cal M}_s = \sigma_{\rm turb}/c_s.
\label{Eq8a}
\end{equation}

The relative errors in $\Delta v$, $\delta_{\Delta v}$,  and in \Tkin, $\delta_{\scriptscriptstyle \rm T}$, 
propagate into the relative error of the Mach number, $\delta_{\scriptscriptstyle \cal M}$, as 
\begin{equation}
\delta_{\scriptscriptstyle \cal M} = \sqrt{ \left( \frac{\delta_{\Delta v}}{8\ln 2}\frac{\Delta v^2}{\sigma^2_{\rm turb}} 
\right)^2 + \left( \frac{ \delta_{\scriptscriptstyle \rm T}}{4}\frac{v^2_{\rm th}}{\sigma^2_{\rm turb}}\right)^2 +
\left( \frac{\delta_{\scriptscriptstyle \rm T} }{2}\right)^2 }, 
\label{Eq08a}
\end{equation}
where $\delta_{\scriptscriptstyle \rm T} \sim 0.2$ and $\delta_{\Delta v} \sim 0.01 - 0.05$ (see Table~\ref{tbl-1}).

The value of the linewidth \Dv\ and the core
radius $R$ can be used to estimate the virial mass (e.g., Lemme \etal\ 1996):
\begin{equation}
M_{\rm vir} = 250 \Delta v^2 R ,
\label{Eq9a}
\end{equation}
where $\Delta v$ is in \kms, $R$ in pc, and $M_{\rm vir}$ in solar masses $M_\odot$.

\clearpage

\clearpage

\clearpage

\clearpage

\clearpage

\clearpage
\begin{table*}
\centering
\caption{Observed parameters of the \nhhh(1,1) and (2,2) lines and calculated model parameters
for \object{SS1} and \object{SS2} } 
\label{tbl-1a}
\begin{tabular}{c r r c c c c c c c c c c}
\hline
\hline
\noalign{\smallskip}
\multicolumn{1}{c}{Peak} &
\multicolumn{1}{c}{$\Delta \alpha$} & 
\multicolumn{1}{c}{$\Delta \delta$} &
\multicolumn{1}{c}{$V_{\scriptscriptstyle\rm LSR}$} & 
\multicolumn{1}{c}{$\Delta v$} &
\multicolumn{1}{c}{\Tmb} & 
\multicolumn{1}{c}{\Tex} & 
\multicolumn{1}{c}{\Trot} & 
\multicolumn{1}{c}{\Tkin} & 
\multicolumn{1}{c}{$n_{\scriptscriptstyle {\rm H}_2}/10^4$} &
\multicolumn{1}{c}{$\tau_{11}$} & 
\multicolumn{1}{c}{$\tau_{22}$} & 
\multicolumn{1}{c}{$N_{\rm tot}/10^{14}$}\\ 
\multicolumn{1}{c}{Id.} &
\multicolumn{1}{c}{$('')$} & 
\multicolumn{1}{c}{$('')$} & 
\multicolumn{1}{c}{(\kms)} &
\multicolumn{1}{c}{(\kms)} & 
\multicolumn{1}{c}{(K)} & 
\multicolumn{1}{c}{(K)} &  
\multicolumn{1}{c}{(K)} & 
\multicolumn{1}{c}{(K)} &
\multicolumn{1}{c}{(\cmm)} & & & 
\multicolumn{1}{c}{(\cm)} \\ 
\noalign{\smallskip}
\hline
\noalign{\smallskip}
\multicolumn{13}{c}{\object{SS1}}\\
\noalign{\smallskip}
$\alpha$  & $-40$ & 0  & 6.14 & 0.54 & 1.1 & 5.5 & 12.5 & 13.6 & 1.8 & 1.4 & 0.1 & 1.8 \\[-1pt]
 & 0 & 0 & 6.41 & 0.69 & 0.9 & 5.8 & 13.6 & 15.1 & 1.9 & 0.9 & 0.09 & 1.4 \\[-1pt]
 & $-40$ & $-40$ & 6.32 & 0.95 & 0.8 & 6.9 & 12.9 & 14.1 & 3.4 & 0.5 & 0.04 & 1.3 \\

\noalign{\smallskip}
\multicolumn{13}{c}{\object{SS2A}}\\
\noalign{\smallskip}
$\alpha$  & $-40$ & 200  & 7.36 & 0.47 & 2.9 & 6.1 & 11.0 & 11.8 & 2.7 & 6.0 & 0.3 & 9.1 \\[-1pt]
$\beta$ & $-80$ & $-40$ & 6.62 & 0.42 & 2.4 & 5.5 & 10.0 & 10.5 & 2.2 & 6.4 & 0.2 & 9.0 \\[-1pt]
$\gamma$ & $-360$ & $320$ & 6.46 & 0.24 & 2.0 & 4.9 & 10.2 & 10.8 & 1.6 & 9.1 & 0.4 & 6.3 \\[-1pt]
$\delta$ & $-240$ & $240$ & 6.78 & 0.30 & 1.9 & 5.0 & 10.8 & 11.5 & 1.6 & 7.0 & 0.3 & 5.7 \\[-1pt]
         & $0$ & $160$ & 7.29 & 0.30 & 2.5 & 6.0 & 11.5 & 12.4 & 2.4 & 4.9 & 0.3 & 4.4 \\[-1pt]
        & $40$ & $120$ & 7.31 & 0.24 & 1.3 & 4.6 & 10.4 & 11.0 & 1.3 & 4.3 & 0.2 & 2.8 \\[-1pt]
        & $40$ & $-40$ & 7.15 & 0.43 & 1.1 & 4.5 & 11.8 & 12.7 & 1.1 & 2.7 & 0.2 & 2.5 \\[-1pt]
       & $-40$ & $0^a$ & 6.44 & 0.26 & 1.5 & 4.0 & 11.5 & 12.3 & 0.7 & 1.5 & 0.09 & 0.8 \\[-1pt]
        &  \multicolumn{2}{c}{}           & 7.25 & 1.37 &  & &  &  & & 0.4 &      &  \\[-1pt]
      & $-40$ & $160$  & 7.26 & 0.54 & 2.1 & 5.3 & 12.9 & 14.2 & 1.6 & 4.2 & 0.4 & 5.3 \\[-1pt]
    & $-80$ & $0^a$    & 6.64 & 0.59 & 1.8 & 3.5 & 9.8 & 10.3 & 4.6 & 1.8 & 0.06 & 2.4 \\[-1pt]
    & \multicolumn{2}{c}{}    & 7.36 & 1.26 &  &  &  &  &  & 0.5 &  &  \\[-1pt]
      & $-80$ & $40^a$  & 6.52 & 0.28 & 1.8 & 3.9 & 10.7 & 11.4 & 0.7 & 2.0 & 0.09 & 1.2 \\[-1pt]
  & \multicolumn{2}{c}{} & 7.44 & 1.43 &  & &  & & & 0.5 &  & \\[-1pt]
      & $-80$ & $200$  & 7.48 & 0.37 & 2.7 & 6.0 & 9.8 & 10.3 & 3.1 & 5.5 & 0.2 & 7.7 \\[-1pt]
      & $160$ & $80$  & 7.49 & 0.22 & 1.1 & 4.3 & 13.3 & 14.7 & 0.8 & 4.4 & 0.4 & 1.7 \\[-1pt]
     & $-280$ & $240$ & 6.69 & 0.28 & 1.8 & 4.7 & 9.8 & 10.3 & 1.4 & 9.6 & 0.3 & 7.9 \\[-1pt]
     & $-280$ & $280$ & 6.68 & 0.24 & 2.1 & 5.0 & 9.6 & 10.1 & 1.7 & 10.4 & 0.3 & 8.1 \\

\noalign{\smallskip}
\multicolumn{13}{c}{\object{SS2B}}\\
\noalign{\smallskip}
$\alpha$  & $0$ & $0$  & 8.19 & 0.26 & 2.3 & 5.4 & 9.7 & 10.2 & 2.2 & 7.7 & 0.2 & 6.9 \\[-1pt]
$\beta$  & $40$ & $40$  & 8.32 & 0.29 & 2.3 & 5.4 & 9.6 & 10.0 & 2.3 & 6.8 & 0.2 & 7.1 \\[-1pt]
$\gamma$ & $240$ & $80$ & 8.18 & 0.43 & 2.2 & 6.3 & 9.9 & 10.4 & 3.5 & 2.9 & 0.1 & 4.9 \\[-1pt]
         & $0$ & $40$   & 8.24 & 0.32 & 1.6 & 4.7 & 10.2 & 10.8 & 1.3 & 5.7 & 0.2 & 5.0 \\[-1pt]
         & $40$ & $0$   & 8.23 & 0.37 & 1.6 & 4.7 & 10.3 & 10.9 & 1.4 & 5.5 & 0.2 & 5.5 \\[-1pt]
         & $80$ & $0$   & 8.04 & 0.34 & 1.3 & 4.3 & 10.3 & 10.8 & 1.0 & 6.7 & 0.3 & 5.8 \\[-1pt]
         & $80$ & $40$  & 8.18 & 0.44 & 1.2 & 5.5 & 11.1 & 11.9 & 2.0 & 1.7 & 0.09 & 2.1 \\[-1pt]
         & $160$ & $0$  & 8.09 & 0.27 & 1.4 & 4.3 & 9.4 & 9.8 & 1.1 & 8.0 & 0.2 & 6.6 \\[-1pt]
       & $280$ & $80^a$ & 8.30 & 0.25 & 1.7 & 3.8 & 8.5 & 8.8 & 0.7 & 1.2 & 0.02 & 1.0 \\[-1pt]
         & \multicolumn{2}{c}{} & 7.95 & 0.33 &  &  &  &  &  & 1.8 &  &  \\

\noalign{\smallskip}
\hline
\noalign{\smallskip}
\multicolumn{13}{l}{{\bf Notes.}\ $^a$Asymmetric profile; parameters of the second component are given
in the second row.    }
\end{tabular}
\end{table*}

\clearpage
\begin{table*}
\centering
\caption{Observed parameters of the \nhhh(1,1) and (2,2) lines and calculated model parameters
for \object{L1251C}
}
\label{tbl-2a}
\begin{tabular}{c r r c c c c c c c c c c}
\hline
\hline
\noalign{\smallskip}
\multicolumn{1}{c}{Peak} &
\multicolumn{1}{c}{$\Delta \alpha$} & \multicolumn{1}{c}{$\Delta \delta$} &
\multicolumn{1}{c}{$V_{\scriptscriptstyle\rm LSR}$} & \multicolumn{1}{c}{$\Delta v$} &
\multicolumn{1}{c}{\Tmb} & \multicolumn{1}{c}{\Tex} & \multicolumn{1}{c}{\Trot} &
\multicolumn{1}{c}{\Tkin} & \multicolumn{1}{c}{$n_{\scriptscriptstyle {\rm H}_2}/10^4$} &
\multicolumn{1}{c}{$\tau_{11}$} & \multicolumn{1}{c}{$\tau_{22}$} &
\multicolumn{1}{c}{$N_{\rm tot}/10^{15}$}\\
\multicolumn{1}{c}{Id.} &
\multicolumn{1}{c}{$('')$} & \multicolumn{1}{c}{$('')$} & \multicolumn{1}{c}{(\kms)} &
\multicolumn{1}{c}{(\kms)} & \multicolumn{1}{c}{(K)} & \multicolumn{1}{c}{(K)} &
\multicolumn{1}{c}{(K)} & \multicolumn{1}{c}{(K)} &
\multicolumn{1}{c}{(\cmm)} & & & \multicolumn{1}{c}{(\cm)} \\
\noalign{\smallskip}
\hline
\noalign{\smallskip}
$\alpha$ & 0& 0  & $-$4.722(3) & 0.290 & 3.9 & 7.3 & 10.0 & 10.5 & 5.9 & 6.4 & 0.2 & 0.8 \\[-1pt]
 & 40& 0 & $-$4.733(5) & 0.28 & 3.5 & 7.0 & 10.5 & 11.1 & 4.4 & 6.2 & 0.3 & 0.7 \\[-1pt]
 & 40& $-$40 & $-$4.826(7) & 0.33 & 3.0 & 6.6 & 9.8 & 10.3 & 4.2 & 5.0 & 0.2 & 0.7 \\[-1pt]
 & 0& $-$40 & $-$4.818(5) & 0.33 & 3.6 & 7.3 & 10.9 & 11.7 & 4.7 & 5.2 & 0.3 & 0.7 \\[-1pt]
 &$-$40& $-$40 & $-$4.805(6) & 0.35 & 3.4 & 6.9 & 11.9 & 12.9 & 3.5 & 5.4 & 0.4 & 0.6 \\[-1pt]
 &$-$40& 0 & $-$4.718(6) & 0.29 & 3.3 & 6.8 & 10.5 & 11.1 & 4.1 & 5.9 & 0.3 & 0.7 \\[-1pt]
&40&40 & $-$4.625(6) & 0.24 & 2.6 & 6.3 & 11.4 & 12.2 & 2.8 & 4.7 & 0.3 & 0.4 \\[-1pt]
 &$-$80&$-$40 & $-$4.797(7) & 0.34 & 3.2 & 7.0 & 11.4 & 12.2 & 3.9 & 4.5 & 0.3 & 0.6 \\[-1pt]
 &80&0 & $-$4.755(6) & 0.29 & 2.9 & 6.4 & 10.4 & 11.0 & 3.4 & 5.4 & 0.2 & 0.6 \\[-1pt]
 &80&$-$40 & $-$4.88(1) & 0.29 & 1.6 & 4.7 & 10.9 & 11.7 & 1.3 & 5.6 & 0.3 & 0.4 \\[-1pt]
 &$-$120&0 & $-$4.728(9) & 0.33 & 2.3 & 5.4 & 12.0 & 12.9 & 1.8 & 6.2 & 0.4 & 0.5 \\[-1pt]

\noalign{\smallskip}
\hline
\end{tabular}
\end{table*}

\end{appendix}

\end{document}